\documentclass[english,pra,,showpacs,reprint]{revtex4-1}
\usepackage[T1]{fontenc}
\usepackage[latin9]{inputenc}
\usepackage{color}
\usepackage{babel}
\usepackage{verbatim}
\usepackage{amsmath}
\usepackage{graphicx}
\usepackage{esint}
\usepackage{xargs}[2008/03/08]
\usepackage[unicode=true,pdfusetitle,
 bookmarks=true,bookmarksnumbered=true,bookmarksopen=true,bookmarksopenlevel=1,
 breaklinks=false,pdfborder={0 0 0},backref=false,colorlinks=true]
 {hyperref}

\makeatletter
\usepackage[caption=false]{subfig}
\makeatletter

\@ifundefined{showcaptionsetup}{}{%
 \PassOptionsToPackage{caption=false}{subfig}}
\usepackage{subfig}
\makeatother

\begin{document}

\title{Suppressing technical noises in weak measurement by entanglement}

\author{Shengshi Pang}

\email{shengshp@usc.edu}

\author{Todd A. Brun}

\email{tbrun@usc.edu}

\affiliation{Department of Electrical Engineering, University of Southern California,
Los Angeles, California 90089, USA}
\begin{abstract}
Postselected weak measurement has aroused broad interest for its distinctive
ability to amplify small physical quantities. However, the low postselection
efficiency to obtain a large weak value has been a big obstacle to
its application in practice, since it may waste resources, and reduce
the measurement precision. An improved protocol was proposed in {[}Phys.
Rev. Lett. \textbf{113}, 030401 (2014){]} to make the postselected
weak measurement dramatically more efficient by using entanglement.
Such a protocol can increase the Fisher information of the measurement
to approximately saturate the well-known Heisenberg limit. In this
paper, we review the entanglement-assisted protocol of postselected
weak measurement in detail, and study its robustness against technical
noises. We focus on readout errors. Readout errors can greatly degrade
the performance of postselected weak measurement, especially when
the readout error probability is comparable to the postselection probability.
We show that entanglement can significantly reduce the two main detrimental
effects of readout errors: inaccuracy in the measurement result, and
the loss of Fisher information. We extend the protocol by introducing
a majority vote scheme to postselection to further compensate for
readout errors. With a proper threshold, almost no Fisher information
will be lost. These results demonstrate the effectiveness of entanglement
in protecting postselected weak measurement against readout errors.
\end{abstract}

\pacs{03.65.Ta, 03.65.Ud, 03.65.Ca, 03.67.Ac}

\maketitle
\global\long\def\vs{\overrightarrow{\sigma}}
\global\long\def\vm{\overrightarrow{m}}
\global\long\def\vx{\overrightarrow{x}}
\global\long\def\vr{\overrightarrow{r}}
\global\long\def\ha{\hat{A}}
\global\long\def\hf{\hat{F}}
\global\long\def\i{\mathrm{i}}
\global\long\def\e{\mathrm{e}}
\global\long\def\d{\mathrm{d}}
\global\long\def\rl{r_{\mathrm{loss}}}
\newcommandx\si[1][usedefault, addprefix=\global, 1=]{|\Psi_{i}^{#1}\rangle}
\global\long\def\re{\mathrm{Re}}
\global\long\def\im{\mathrm{Im}}
\global\long\def\tr{\mathrm{tr}}
\global\long\def\var{\mathrm{Var}}
\global\long\def\et{\e^{\i\theta}}
\newcommandx\sfi[1][usedefault, addprefix=\global, 1=]{\langle\Psi_{f}^{#1}|\Psi_{i}^{#1}\rangle}
\global\long\def\rdf{\rho_{D}^{\prime}}
\newcommandx\df[1][usedefault, addprefix=\global, 1=]{|D_{f#1}\rangle}
\global\long\def\di{|D\rangle}
\newcommandx\csf[1][usedefault, addprefix=\global, 1=]{\langle\Psi_{f}^{#1}|}
\newcommandx\aw[1][usedefault, addprefix=\global, 1=]{\frac{\langle\Psi_{f}^{#1}|\ha|\Psi_{i}^{#1}\rangle}{\langle\Psi_{f}^{#1}|\Psi_{i}^{#1}\rangle}}
\global\long\def\hm{\hat{M}}
\global\long\def\net{\e^{-\i\theta}}
\newcommandx\csi[1][usedefault, addprefix=\global, 1=]{\langle\Psi_{i}^{#1}|}
\newcommandx\ps[1][usedefault, addprefix=\global, 1=]{P_{s}^{(#1)}}
\newcommandx\sff[1][usedefault, addprefix=\global, 1=]{|\Psi_{f}^{#1}\rangle}
\newcommandx\av[1][usedefault, addprefix=\global, 1={\rm total}]{\ha_{#1}}
\global\long\def\ko{|0\rangle}
\global\long\def\kon{\ko^{\otimes n}}
\global\long\def\kl{|1\rangle}
\global\long\def\kln{\kl^{\otimes n}}
\newcommandx\qlo[1][usedefault, addprefix=\global, 1=]{q_{1\rightarrow0}^{#1}}
\newcommandx\qol[1][usedefault, addprefix=\global, 1=]{q_{0\rightarrow1}^{#1}}
\global\long\def\pe{p(\mathrm{error}|\kon)}
\global\long\def\cdi{\langle D|}
\global\long\def\dav#1{\langle D|#1|D\rangle}
\global\long\def\peo{p(\mathrm{error}|\ko)}
\global\long\def\hv{\hat{V}}
\global\long\def\hu{\hat{U}}
\global\long\def\hrx{\hat{R}_{x}}
\global\long\def\hrz{\hat{R}_{z}}
\global\long\def\vp{\mathcal{V}^{\perp}}
\newcommandx\ig[1][usedefault, addprefix=\global, 1=]{I_{g}^{#1}}
\global\long\def\iq{I_{g}^{(Q)}}
\global\long\def\fg{\Phi_{g}}
\global\long\def\pg{\partial_{g}\fg}
\global\long\def\bk#1#2{\langle#1|#2\rangle}
\global\long\def\ss#1{|\psi^{#1}\rangle}
\global\long\def\ket#1{|#1\rangle}
\global\long\def\avdf#1#2{\langle D_{f}^{#1}|#2|D_{f}^{#1}\rangle}
\global\long\def\avd#1{\langle D|#1|D\rangle}
\global\long\def\sx{\hat{\sigma}_{x}}
\global\long\def\pv{\partial_{\varphi}}
\global\long\def\iv{I_{\varphi}}
\global\long\def\hp{\hat{p}}
\global\long\def\avg#1{\langle\fg|#1|\fg\rangle}
\global\long\def\pon{p_{0}^{(n)}}
\global\long\def\pln{p_{1}^{(n)}}
\global\long\def\pol{p_{0}^{(1)}}
\global\long\def\pll{p_{1}^{(1)}}
\global\long\def\wo{w_{0,j}}
\global\long\def\wl{w_{1,j}}
\global\long\def\wok{w_{0,j}^{(k)}}
\global\long\def\wlk{w_{1,j}^{(k)}}
\global\long\def\ivo{I_{\varphi,0}}
\global\long\def\hk{h_{j}^{(k)}}
\global\long\def\sk{\sum_{j=0}^{k}\binom{n}{j}}
\global\long\def\rlp{r_{\mathrm{loss}}^{\prime}}
\global\long\def\sz{\hat{\sigma}_{z}}
\global\long\def\eon{\eta_{0}^{(n)}}
\global\long\def\eln{\eta_{1}^{(n)}}
\global\long\def\gn{\gamma(n)}
\global\long\def\gnk{\gamma(n,k)}
\global\long\def\he{\hat{E}}

\section{Introduction}

A quantum measurement is associated with an observable of the system
in the standard von Neumann model. The effect of a quantum measurement
is to stochastically project the system onto an eigenstate of the
observable, and the reading from the measurement is the corresponding
eigenvalue. Such a projective measurement usually requires the interaction
between the system and the pointer to be strong, or the spread of
the pointer wavefunction to be narrow, so that the final wavefunctions,
translated by different eigenvalues of the observable, can be distinguished
accurately.

In 1988, Aharonov, Albert, and Vaidman (AAV) \cite{Aharonov1988}
coined a quantum measurement protocol that violated the above conditions
for projective measurements. In this protocol, the width of the pointer
wave function is larger than the eigenvalue separation, or equivalently
the interaction between the system and the pointer is weak, so that
the final states of the pointer, kicked by different eigenvalues,
have very large overlap with each other. AAV also introduced postselection
of the system by measuring the system as well as the pointer, and
retaining only those events where the system measurement produced
a particular value. Such postselection destroys the correlation between
the system and the pointer, and collapses the pointer to interfere
the overlapping translations.

The AAV protocol of weak measurement with postselection seems trivial,
but it can have a surprising physical effect: with a proper postselection
of the system, the average translation of the pointer can be much
larger than any eigenvalue of the observable, in sharp contrast to
the standard projective quantum measurement. The mechanism behind
this effect is that with the postselection of the system, the interference
between the component wavefunctions, translated by different eigenvalues
in the pointer state, may dramatically cancel the major part of the
original wavefunction, resulting in a shift of the pointer that goes
far beyond the eigenvalue spectrum of the observable.

The large shift of the pointer can be approximated as a linear amplification
of the otherwise weak interaction strength. Such a linear amplification
can be characterized by a quantity called the \emph{weak value }\cite{Aharonov1988}
(which will be introduced in detail in Sec. \ref{sec:Preliminary}).
This value can have formally divergent behavior when the postselected
state of the system asymptotically becomes orthogonal to the initial
state of the system. (Of course, in practice, the amplification effect
cannot be infinitely large. This has been extensively investigated
in recent years \cite{Koike2011,Susa2012,DiLorenzo2014,Pang2014}.)

There was some controversy over the condition for the validity of
the weak value after the birth of AAV weak measurement, but it was
soon clarified \cite{Duck1989}. In addition, unlike the eigenvalues
of an observable, the weak value is generally a complex value, with
its real and imaginary parts playing different roles in the amplification
effect \cite{Jozsa2007}.

The measurement of weak values has been realized experimentally \cite{Pryde2005},
and the amplification effect has been found useful in observing weak
physical effects in many real experiments, including the spin Hall
effect of light \cite{Hosten2008,Gorodetski2012,Zhou2012}, optical
beam deflection \cite{Ritchie1991,Dixon2009,Starling2009,Pfeifer2011,Turner2011,Viza2014,Goswami2014},
optical frequency shift \cite{Starling2010}, optical phase shift
\cite{Starling2010a,Xu2013}, temperature shift \cite{Egan2012},
temporal shift \cite{Strubi2013,Viza2013,Mirhosseini2014}, etc. More
experimental protocols have also been proposed \cite{Brunner2010,Feizpour2011,Li2011,Zilberberg2011,Gotte2012,Nishizawa2012,Wu2012,Dressel2013,Hayat2013,Strubi2013,Zhou2013,Huang2015}.
Moreover, weak measurements have been realized on systems other than
optical systems, including superconducting circuits \cite{Palacios-Laloy2010,Groen2013,Campagne-Ibarcq2014},
NMR \cite{Lu2014}, among others. For a comprehensive review of postselected
weak measurement and weak value amplification, we refer the readers
to \cite{Shikano2011,Kofman2012,Dressel2014,Dressel2015}.

When postselected weak measurement is used to amplify small physical
quantities, a large weak value is usually desired. However, this will
lead to a low postselection efficiency, since the weak value is approximately
reciprocal to the square root of the postselection probability. Low
postselection efficiency may result in a waste of resources and reduce
the Fisher information of the measurement if the failed postselections
are discarded, and consequently cancel the advantage of weak value
amplification. In fact, this has led to a recent hot debate: whether
postselecting the system and discarding the unselected events can
ever increase the precision of the measurement.

Some studies have suggested that the weak value amplification can
produce higher precision in weak measurements \cite{Starling2009,Feizpour2011,Nishizawa2012,Hayat2013},
while others found the opposite results \cite{Zhu2011,Knee2013,Nishizawa2015}.
More extensive work on this issue has made clear that the Fisher information
of weak measurements cannot be increased by postselection if the unselected
events are discarded \cite{Knee2013,Ferrie2014a,Combes2014}, because
the discarded events take away some Fisher information \cite{Tanaka2013},
and the distribution of postselection probabilities can also carry
information \cite{Ferrie2014a,Combes2014,Zhang2015}. Interestingly,
however, if all these sources of Fisher information are taken into
account, the total Fisher information can saturate the Heisenberg
limit in some cases, even with seemingly classical resources \cite{Zhang2015,Jordan2015b}.
Moreover, weak value amplification can give an advantage in suppressing
technical noise \cite{Jordan2014} (although not all types of noises
can be suppressed \cite{Knee2014a}), or even use technical noises
to enhance the sensitivity \cite{Kedem2012}. A brief review of the
recent controversy over weak value amplification can be found in \cite{Knee2014}.

Because of the above problems of low postselection efficiency, an
important goal in the practice of weak value amplification is to raise
the postselection probability. Some efforts have been made on this.
For example, it was proposed that by recycling unpostselected photons,
almost every photon can eventually be successfully postselected \cite{Dressel2013}.

In a recent study \cite{Pang2014a}, it was noticed that for a given
weak value, the choice of pre- and postselection of the system to
realize it is usually not unique, so there is some freedom to maximize
the postselection probability. Alternatively, if the postselection
probability is fixed, there is some freedom to maximize the magnitude
of the weak value. Maximizing the postselection probability or weak
value can dramatically improve the resource usage or performance of
the weak value amplification. And the result of either optimization
shows that the loss of Fisher information in the discarded events
can be made as small as the order of the interaction strength, which
is usually negligible in the weak value approximation. This implies
that postselected weak measurement can offer technical advantages
\cite{Jordan2014} at almost no cost in Fisher information.

Based on these optimization results for the postselection efficiency
and weak value, an improved protocol of postselected weak measurement
assisted by entanglement was proposed in \cite{Pang2014a}. The protocol
uses entangled systems rather than uncorrelated systems. If there
are $n$ systems entangled initially, the postselection probability
scales as $n^{2}$, while if the $n$ systems are uncorrelated, the
postselection probability can only scale linearly in $n$. Thus, entanglement
between the systems can bring a marked increase (of order $n$) in
the postselection probability. An important consequence of this increase
in postselection probability is that the Fisher information can also
be correspondingly raised by the order of $n$ (because the Fisher
information is proportional to the size of the data sample), and approach
the Heisenberg limit, which is the upper bound on estimation precision
achievable in quantum metrology \cite{Giovannetti2006}.

This paper builds on \cite{Pang2014a} to detail the entanglement-assisted
protocol of weak measurement, and its advantages in improving the
metrological performance of weak measurement. Furthermore, since technical
noise is inevitable in real experimental devices, we will study the
influence of technical noise on this protocol, and show the robustness
of this protocol against the noise. The main technical noise we will
consider is readout error in postselecting the system.

Readout errors mix successful postselection events with unsuccessful
ones. Since the shift of the pointer resulting from unsuccessful postselections
is much smaller than from successful postselections, and the sensitivity
of the pointer states in the former case is also much lower, mixing
them will bring errors to the measurement result and reduce the precision
of the measurement. Moreover, as will be shown, readout errors may
be more detrimental in postselected weak measurements than in other
quantum measurements. When the postselection probability is very small,
even a low rate of readout errors may cause severe problems. So it
is necessary to suppress the influence of readout errors to make postselected
weak measurements more reliable in practice.

In this paper, we will analyze the robustness of postselected weak
measurement against readout errors, and show that the use of entanglement
can correct the deviation of measurement result caused by readout
errors with an extremely high success rate. Furthermore, it will be
shown that entanglement can recover part of the Fisher information
reduced by readout errors. Introducing an appropriate measurement
threshold strategy can decrease the Fisher information loss to nearly
zero. These results suggest that entanglement, combined with a measurement
threshold strategy, can effectively suppress the effect of readout
errors on postselected weak measurements.

This paper is organized as follows. First, in Sec. \ref{sec:Preliminary},
we briefly introduce the weak value theory of postselected weak measurement
and how it leads to the amplification effect. Then, in Sec. \ref{sec:Optimization},
we study the optimization of postselection to maximize the postselection
probability given the weak value, or to maximize the weak value given
the postselection probability. Sec. \ref{sec:Entanglement-assisted}
is devoted to introducing entanglement-assisted weak measurement based
on the optimization result. Sec. \ref{sec:Fisher-information-of}
gives a detailed study of the Fisher information of postselected weak
measurement, and shows that the Fisher information can saturate the
Heisenberg limit with the assistance of entanglement. A qubit example
is given in Sec. \ref{sec:Qubit-example} to illustrate the entanglement-assisted
protocol, and verify the saturation of Heisenberg limit. Finally,
in Sec. \ref{sec:Readout-errors}, we investigate the influence of
readout errors in detail, and introduce a measurement threshold scheme
to protect the Fisher information against readout errors.

\section{Review of weak value formalism\label{sec:Preliminary}}

In a standard quantum measurement, the measurement results are eigenvalues
of a system observable, and the system collapses to the eigenstate
of the observable corresponding to the measurement result. A typical
model to realize this standard quantum measurement is
\begin{equation}
H_{\mathrm{int}}=g\ha\otimes\hf\delta(t-t_{0}),\label{eq:12}
\end{equation}
where $\ha$ and $\hf$ are observables of the system and the pointer
respectively, and $g$ characterizes the strength of the interaction.
Suppose the initial state of the system is $\si$, and and initial
state of the pointer is $\di$, then the system and the pointer are
coupled by the interaction, and evolve to an entangled state
\begin{equation}
|\Phi\rangle=\exp(-\i g\ha\otimes\hf)\si\di.
\end{equation}
If we expand $\si$ along the eigenstates of $\ha$, $|\Phi\rangle$
can be written as
\begin{equation}
|\Phi\rangle=\sum_{k}c_{k}|a_{k}\rangle\exp(-\i ga_{k}\hf)\di,\label{eq:19}
\end{equation}
where $a_{k},\,|a_{k}\rangle$ are eigenvalues and eigenstates of
$\ha$, and $c_{k}$ are the expansion coefficients of $\si$ in the
basis of $\{|a_{k}\rangle\}$.

Different $\exp(-\i ga_{k}\hf)$ in (\ref{eq:19}) transform $\di$
into different states. If $\di$ is properly chosen so that the overlaps
between $\exp(-\i ga_{k}\hf)\di$ are sufficiently small, the $\exp(-\i ga_{k}\hf)\di$
can be distinguished with a low error probability then, and the measurement
on the pointer will make the system collapse to a state that is close
to an eigenstate of $\ha$. For example, suppose $\hf$ is the momentum
operator $\hp$, then $(-\i ga_{k}\hp)$ is a translation operator
in the position space of the pointer, and if one can measure the position
of the pointer, the measurement results will be $ga_{k}$, and the
system will collapse to $|a_{k}\rangle$. If $g$ is set to $1$,
it will lead to the formalism of standard projective quantum measurement.

A major innovation by AAV in weak measurements was introducing postselection
of the system. This seemingly minor change turns out to give some
surprising results that are dramatically different from ordinary quantum
measurements. 

In detail: if the system is postselected to the state $\sff$ after
the interaction, the pointer then collapses to the (unnormalized)
state
\begin{equation}
|D_{f}\rangle=\csf\exp(-\i g\ha\otimes\hf)\si\di.
\end{equation}

When $g$ is sufficiently small, the $\df$ is approximately
\begin{equation}
\begin{aligned}\df & \approx\csf(1-\i g\ha\otimes\hf)\si\di\\
 & =\sfi(1-\i gA_{w}\hf)\di,
\end{aligned}
\label{eq:11}
\end{equation}
where $A_{w}$ is the weak value,
\begin{equation}
A_{w}=\aw.\label{eq:8}
\end{equation}
If $g$ is so small that $gA_{w}\ll1$, $\df$ can be rewritten as
\begin{equation}
\df\approx\exp(-\i gA_{w}\hf)\di.
\end{equation}
So one can see that in the presence of postselection of the system,
the pointer is shifted by roughly $gA_{w}$ (in the representation
of a complementary observable of $\hf$). In sharp contrast to ordinary
measurement, the shift of the pointer in this case can be much larger
than any $ga_{k}$, because $A_{w}$ can be much larger than $1$
when $|\sfi|\ll1$. 

Note that $A_{w}$ can be complex, and in this case $\exp(-\i gA_{w}\hf)$
is not just a simple translation operator. In fact, it can be decomposed
to the product of a translation operator (corresponding to the real
part of $A_{w}$) and a state reduction operator (corresponding to
the imaginary part of $A_{w}$). Jozsa gave a very detailed study
of complex weak value in \cite{Jozsa2007}, and analyzed the role
of the real and imaginary parts of the weak value. He showed that
if the pointer observable $\hf$ is the momentum $\hat{p}$, then
the shifts in the average position and momentum of the pointer are,
respectively,
\begin{equation}
\begin{aligned}\langle\Delta\hat{q}\rangle= & g\mathrm{Re}A_{w}+g\mathrm{Im}A_{w}(m\frac{\mathrm{d}}{\mathrm{d}t}\mathrm{Var}_{\hat{q}}),\\
\langle\Delta\hat{p}\rangle= & 2g\mathrm{Im}A_{w}\mathrm{Var}_{\hat{p}},
\end{aligned}
\label{eq:33}
\end{equation}
where $\hat{q}$ and $\hat{p}$ are the position and momentum operators
of the pointer.

That result can be generalized to a more general form. Suppose one
measures an observable $\hm$ on the pointer after postselecting the
system. The average shift of the pointer is
\begin{equation}
\langle\Delta\hm\rangle_{f}=\frac{\langle D_{f}|\hm|D_{f}\rangle}{\langle D_{f}|D_{f}\rangle}-\langle\hm\rangle_{\di}.\label{eq:6}
\end{equation}
From Eq. (\ref{eq:11}), one can get
\begin{equation}
\begin{aligned}\langle D_{f}|\hm|D_{f}\rangle & \approx|\sfi|^{2}(\langle\hm\rangle_{\di}+\i g\re A_{w}\langle[\hf,\hm]\rangle_{\di}\\
 & +g\im A_{w}\langle\{\hf,\hm\}\rangle_{\di}),\\
\langle D_{f}|D_{f}\rangle & \approx|\sfi|^{2}(1+2g\im A_{w}\langle\hf\rangle_{\di}),
\end{aligned}
\label{eq:3-1}
\end{equation}
so plugging (\ref{eq:3-1}) into (\ref{eq:6}) produces
\begin{equation}
\begin{aligned}\langle\Delta\hm\rangle_{f} & \approx g\im A_{w}(\langle\{\hf,\hm\}\rangle_{\di}-2\langle\hf\rangle_{\di}\langle\hm\rangle_{\di})\\
 & +\i g\re A_{w}\langle[\hf,\hm]\rangle_{\di}.
\end{aligned}
\label{eq:7}
\end{equation}
Note that if one plugs $\hf=\hat{p}$ and $\hm=\hat{q},\,\hat{p}$
into (\ref{eq:7}), the result (\ref{eq:33}) can be immediately recovered.

Eqs. (\ref{eq:33}) and (\ref{eq:7}) imply that the shift of the
pointer is roughly proportional to the weak value $A_{w}$ when $g\ll1$.
Since $A_{w}$ can be much larger than $1$ when $\sfi\ll1$, the
shift of the pointer can be treated as an amplification of $g$ by
the weak value $A_{w}$. This is the origin of the amplification effect
in postselected weak measurements. This amplification effect has been
widely used in experiments to measure small parameters, as reviewed
in the introduction.

\section{Optimization of weak measurements\label{sec:Optimization}}

A shortcoming of postselected weak measurements that can be seen immediately
from (\ref{eq:8}) is that a large weak value $A_{w}$ requires a
very small overlap between the initial state and the postselected
state of the system. This implies that the postselection probability
must be very low, because when $g\ll1$, the success probability of
a postselection is approximately
\begin{equation}
P_{s}\approx|\sfi|^{2}.\label{eq:post prob}
\end{equation}
Therefore, an important problem in a practical application of weak
measurement amplification is to improve the postselection probability
as much as possible while the weak value is still kept large.

From the definition of the weak value (\ref{eq:8}), when a weak value
$A_{w}$ is fixed, the possible choices of the system initial state
$\si$ and postselected state $\sff$ to realize this weak value is
not unique, and different $\si$ and $\sff$ may give different postselection
probabilities. This provides the possibility to optimize the choice
of pre- and postselections of the system to maximize the the postselection
efficiency for a given weak value. 

Alternatively, if the postselection probability (\ref{eq:post prob})
is given, different pre- and postselections of the system can also
produce different weak values, which will lead to different amplification
abilities for the parameter $g$. So there exist optimal pre- and
postselections of the system to produce the maximum weak value for
a given postselection probability. 

The significance of optimizing weak measurements to maximize either
the postselection probability or the weak value is obvious: one can
reduce the resources needed to give a desired amplification effect,
or one can make the best use of the given resources to produce the
maximum amplification effect. So it is useful for practical applications
of weak value amplification to optimize the performance.

In this section, we study these two ways of optimizing weak measurements
in detail. Besides the significance mentioned above, the optimizations
derived in this section will also be the foundation of the entanglement-assisted
weak measurements that we will introduce in the next section.

\subsection{Maximum postselection probability given the weak value\label{sub:Maximum-postselection}}

In this subsection, we study the first optimization problem introduced
above: that is, to maximize the postselection probability over all
possible pre- and post-selections of the system for a given weak value. 

To solve this problem directly using the constraint on the fixed weak
value $A_{w}$ in maximizing the postselection probability is rather
difficult; so in order to utilize this condition, we first convert
it to another more accessible form. Note that Eq. (\ref{eq:8}) can
be rewritten as
\begin{equation}
\csf(\ha-A_{w})\si=0,
\end{equation}
so the constraint of fixed $A_{w}$ can be reinterpreted as $\sff$
being orthogonal to $(\ha-A_{w})\si$. With this new form of the condition,
the optimization of $|\sfi|^{2}$ can be much simplified.

From a geometrical point of view, it is not difficult to verify that
$\sff$ should be parallel to the component of $\si$ in the subspace
orthogonal to $(\ha-A_{w})\si$ (which we denote as $\vp$ below)
when $|\sfi|^{2}$ is maximized. Therefore, we can first decompose
the initial state of the system along $(\ha-A_{w})\si$ and its orthogonal
subspace $\vp$,
\begin{equation}
\begin{aligned}\si & =\frac{(\ha-A_{w})\si\csi(\ha-A_{w}^{*})\si}{\csi(\ha-A_{w}^{*})(\ha-A_{w})\si}\\
 & +\bigg(\si-\frac{(\ha-A_{w})\si\csi(\ha-A_{w}^{*})\si}{\csi(\ha-A_{w}^{*})(\ha-A_{w})\si}\bigg),
\end{aligned}
\end{equation}
and then the optimal $\sff$ can be obtained:
\begin{equation}
\sff\propto\si-\frac{(\ha-A_{w})\si\csi(\ha-A_{w}^{*})\si}{\csi(\ha-A_{w}^{*})(\ha-A_{w})\si}.\label{eq:11-1}
\end{equation}
Hence, the maximum postselection probability for the given weak value
$A_{w}$ is
\begin{equation}
\begin{aligned}\max P_{s} & =\bigg\|\si-\frac{(\ha-A_{w})\si\csi(\ha-A_{w}^{*})\si}{\csi(\ha-A_{w}^{*})(\ha-A_{w})\si}\bigg\|^{2}\\
 & =\frac{\text{Var}(\ha)_{\si}}{\csi\ha^{2}\si-2\csi\ha\si\re A_{w}+|A_{w}|^{2}},
\end{aligned}
\label{eq:12-1}
\end{equation}
where $\text{Var}(\ha)_{\si}=\csi\ha^{2}\si-\csi\ha\si^{2}$ is the
variance of $\ha$ in the state $\si$.

For the purposes of weak value amplification, we usually desire $|A_{w}|$
to be larger than any eigenvalue of $\ha$, $|A_{w}|\gg\max|\lambda(\ha)|$,
which implies that
\begin{equation}
\begin{aligned}|A_{w}| & \gg\csi\ha\si,\\
|A_{w}| & \gg\sqrt{\csi\ha^{2}\si}.
\end{aligned}
\end{equation}
Therefore, the maximum $P_{s}$ can be approximated as
\begin{equation}
\max P_{s}\approx\frac{\text{Var}(\ha)_{\si}}{|A_{w}|^{2}},\label{eq:9}
\end{equation}
for a large $A_{w}$.

\subsection{Maximum weak value given the postselection probability\label{sub:Maximum-weak-value}}

In this subsection, we solve the second optimization problem that
was introduced at the beginning of this section: that is, to maximize
the weak value over all possible pre- and postselections of the system
for a given postselection probability. Of course, the weak value is
generally complex, so we will focus on maximizing the amplitude of
the weak value.

Suppose the postselection probability is $P_{s}$. Since the phase
of the post-selected state $\sff$ does not affect the weak value
$A_{w}$, $\sff$ can be written as
\begin{equation}
\sff=\sqrt{P_{s}}\si+\sqrt{1-P_{s}}\si[\perp],\label{eq:28}
\end{equation}
where $\si[\perp]$ is a state orthogonal to $\si$. This implies
that we can write the weak value in Eq. (\ref{eq:8}) as 
\begin{equation}
A_{w}=\csi\ha\si+\sqrt{\frac{1-P_{s}}{P_{s}}}\csi[\perp]\ha\si.
\end{equation}
Now to maximize $A_{w}$ is just to maximize $\csi[\perp]\ha\si$
over $\si[\perp]$. 

Similar to the maximization procedure in the last subsection, the
maximum $\csi[\perp]\ha\si$ can be achieved when $\si[\perp]$ is
parallel to the component of $\ha\si$ in the complementary subspace
orthogonal to $\si$, so
\begin{equation}
\begin{aligned}|\csi[\perp]\ha\si|_{\max} & =\big\|\ha\si-\si\csi\ha\si\big\|\\
 & =\sqrt{\text{Var}(\ha)_{\si}}.
\end{aligned}
\end{equation}
Therefore, the largest weak value that can be obtained from the initial
state $\si$ with a given post-selection probability $P_{s}$ is
\begin{equation}
\max|A_{w}|=\csi\ha\si+\sqrt{\frac{1-P_{s}}{P_{s}}\,\text{Var}(\ha)_{\si}}.\label{eq:1-1}
\end{equation}
For a large $|A_{w}|$, $P_{s}\ll1$, so the first term in (\ref{eq:1-1})
can be neglected, thus,
\begin{equation}
\max|A_{w}|\approx\sqrt{\frac{\text{Var}(\ha)_{\si}}{P_{s}}}.\label{eq:26-1}
\end{equation}

The results of both optimization problems, Eqs. (\ref{eq:9}) and
(\ref{eq:26-1}), indicate that the maximum postselection probability
or the maximum weak value are proportional to the variance (or the
square root of the variance) of the observable $\ha$ in the initial
state of the system. This observation leads directly to the entanglement-assisted
weak measurement protocol that we introduce in the next section.

\section{Entanglement-assisted weak measurement\label{sec:Entanglement-assisted}}

As mentioned at the end of the last section, both the maximum postselection
probability (\ref{eq:9}) and the maximum weak value (\ref{eq:26-1})
are proportional to the variance of $\ha$ (or its square root) in
the initial state of the system. Since the variance of an observable
scales differently with the number of the systems in an entangled
state than in an uncorrelated state, this observation leads to a new
weak measurement protocol that we introduce in this section. We shall
see how entanglement can assist the weak measurement either to save
resources or to improve the amplification.

\subsection{Uncorrelated systems\label{sub:Uncorrelated-systems}}

As a reference example, we first consider the case where the systems
have no correlation between each other initially. Suppose we have
$n$ systems. If these $n$ systems are initially in a product state
$\si[1]\otimes\cdots\otimes\si[n]$, and the postselections are $\sff[1],\cdots,\sff[n]$,
which give the same weak value to each individual system, i.e.,
\begin{equation}
\aw[k]=A_{w},\,k=1,\cdots,n,
\end{equation}
then, when $|\sfi[k]|\ll1$, the probability to have at least one
successful event in postselecting these $n$ systems is
\begin{equation}
\begin{aligned}\ps[n] & =1-\prod_{k=1}^{n}\big(1-|\sfi[k]|^{2}\big)\\
 & \approx\sum_{k=1}^{n}|\sfi[k]|^{2}.
\end{aligned}
\end{equation}
Now, if the choice of pre- and postselections for each system maximizes
$|\sfi[k]|^{2}$, i.e., $|\sfi[k]|^{2}=\ps[1]$, where $\ps[1]$ is
the maximal postselection probability for a single system, then
\begin{equation}
\ps[n]\approx n\ps[1].\label{eq:10}
\end{equation}
This implies that the postselection efficiency increases linearly
with $n$ when the systems are initially uncorrelated.

In fact, the linear scaling of $\ps[n]$ with $n$ in (\ref{eq:10})
is the best scaling that can be obtained with initially uncorrelated
systems. When $\si$ is a product state of $n$ systems, say $\si[(n)]=\si[1]\otimes\cdots\otimes\si[n]$,
it can be verified that
\begin{equation}
\var(\ha^{(n)})_{\si[(n)]}=\var(\ha_{1})_{\si[1]}+\cdots+\var(\ha_{n})_{\si[n]}.
\end{equation}
When each $\si[k]$ maximizes $\var(\ha_{k})$, then
\begin{equation}
\var(\ha^{(n)})_{\si[(n)]}=n\var(\ha)_{\si},\label{eq:38}
\end{equation}
where we omitted the subscript in $\ha$ and the superscript in $\ss{}$
on the right side of (\ref{eq:38}) since each individual system has
the same $\ha$ and $\ss{}$. According to Eq. (\ref{eq:9}), the
maximum postselection probability is proportional to the variance
of $\ha$ in the initial state of the system, so (\ref{eq:38}) implies
that if the initial state of the $n$ systems is a product state,
the postselection probability $\ps[n]$ at most can scale linearly
with $n$.

\subsection{Entangled systems\label{sub:Entangled-systems}}

Now let us remove the constraint that the systems are uncorrelated,
and see whether the postselection probability can be improved. For
$n$ systems, we first need to generalize the observable $\ha$. The
observable $\ha$ becomes the sum of $n$ single-system observables
in this case,
\begin{equation}
\ha^{(n)}=\ha_{1}+\cdots\ha_{n},\label{eq:35}
\end{equation}
where we use the superscript $(n)$ to denote the $n$-system observable
explicitly.

Now suppose we have $n$ systems, and they are initially prepared
in the following entangled state:
\begin{equation}
\si[(n)]=\alpha\ket{a_{\max}}^{\otimes n}+\beta\ket{a_{\min}}^{\otimes n}.\label{eq:29}
\end{equation}
Then it can be obtained directly that
\begin{equation}
\begin{aligned}\var(\ha^{(n)})_{\si[(n)]} & =n^{2}[|\alpha|^{2}a_{\max}^{2}+|\beta|^{2}a_{\min}^{2}\\
 & -(|\alpha|^{2}a_{\max}+|\beta|^{2}a_{\min})^{2}].
\end{aligned}
\end{equation}
One can immediately see that with the entangled state (\ref{eq:29})
(and $\alpha\neq0,\,\beta\neq0$), the scaling of $\var(\ha)_{\si}$
becomes quadratic with $n$, which is higher than with a product $\si$
by order $n$. So the maximum postselection probability can be increased
by order $n$ in this case.

Is quadratic scaling with $n$ optimal when entanglement is used?
And what entangled state of the system maximizes the factor before
$n^{2}$ in the variance $\var(\ha)_{\si}$? To answer these two questions,
let us recall that for an arbitrary Hermitian operator $\hat{\Xi}$,
its maximum variance over all possible states $\ss{}$ is
\begin{equation}
\max_{|\psi\rangle}\var(\hat{\Xi})_{|\psi\rangle}=\frac{1}{4}(\xi_{\max}-\xi_{\min})^{2},\label{eq:27-1}
\end{equation}
where $\xi_{\max}$ and $\xi_{\min}$ are the maximum and minimum
eigenvalues of $\hat{\Xi}$ respectively, and the maximum variance
is attained when
\begin{equation}
|\psi\rangle=\frac{1}{\sqrt{2}}(|\xi_{\max}\rangle+\et|\xi_{\min}\rangle),\label{eq:14-1-1}
\end{equation}
where $|\xi_{\max}\rangle$ and $|\xi_{\min}\rangle$ are the corresponding
eigenstates of $\hat{\Xi}$, and $\et$ is an arbitrary phase. 

Applying this fact to the observable $\ha^{(n)}$ in (\ref{eq:35}),
we obtain
\begin{equation}
\max\var(\ha^{(n)})_{\si}=n^{2}\max\var(\ha),\label{eq:39}
\end{equation}
and
\begin{equation}
\max\var(\ha)=\frac{1}{4}(a_{\max}-a_{\min})^{2}.
\end{equation}
This follows because
\begin{equation}
\lambda_{\max}(\ha^{(n)})=na_{\max},\,\lambda_{\min}(\ha^{(n)})=na_{\min},\label{eq:37}
\end{equation}
where
\begin{equation}
\lambda_{\max}(\ha)=a_{\max},\,\lambda_{\min}(\ha)=a_{\min}.
\end{equation}

From Eq. (\ref{eq:9}), one sees that the maximum postselection probability
over all pre- and postselections of the $n$ systems is
\begin{equation}
\ps[n]\approx n^{2}\ps[1],
\end{equation}
which is increased by order $n$ compared to the uncorrelated state
(\ref{eq:10}).

What price do we pay for such an improvement of postselection probability?
Note that the eigenstates of $\ha^{(n)}$ with eigenvalues (\ref{eq:37})
are
\begin{equation}
|\lambda_{\max}\rangle=|a_{\max}\rangle^{\otimes n},\,|\lambda_{\min}\rangle=|a_{\min}\rangle^{\otimes n}.
\end{equation}
So according to (\ref{eq:14-1-1}), the initial state of the $n$
systems should be
\begin{equation}
\si[(n)]=\frac{1}{\sqrt{2}}(|a_{\max}\rangle^{\otimes n}+\et|a_{\min}\rangle^{\otimes n}),\label{eq:14-1}
\end{equation}
which is an entangled state. Therefore, to improve the postselection
efficiency we need entanglement in the initial state of the $n$ systems.

To obtain the maximum postselection probability, the postselected
state of the $n$ systems should also be carefully chosen. According
to (\ref{eq:11-1}) and (\ref{eq:14-1}), in order to maximize the
postselection probability, the system should be postselected to the
following state after the weak interaction:
\begin{equation}
\begin{aligned}\sff[(n)] & \propto-(na_{\min}-A_{w}^{*})|a_{\max}\rangle^{\otimes n}\\
 & +\et(na_{\max}-A_{w}^{*})|a_{\min}\rangle^{\otimes n}.
\end{aligned}
\label{eq:16}
\end{equation}
When $|A_{w}|\gg\max\{n|a_{\max}|,n|a_{\min}|\}$, $\sff[(n)]$ can
be simplified to
\begin{equation}
\sff[(n)]\propto\e^{\frac{n\Delta}{A_{w}^{*}}}|a_{\max}\rangle^{\otimes n}-\et|a_{\min}\rangle^{\otimes n},
\end{equation}
where $\Delta=a_{\max}-a_{\min}$.

Similarly, we can also use the maximally entangled state (\ref{eq:14-1})
as the initial state of the $n$ systems to increase the weak value
with the postselection probability fixed. The only difference from
the previous protocol is the choice of the postselected state of the
systems. When the postselection probability is fixed to $P_{s}$,
the postselected state contains two components: one is $\sqrt{P_{s}}\si[(n)]$,
and the other is $\sqrt{1-P_{s}}\si[(n)\perp]$. According to Sec.
\ref{sub:Maximum-weak-value}, for the optimal postselection, $\si[(n)\perp]$
should be proportional to the component of $\ha^{(n)}\si[(n)]$ in
the subspace orthogonal to the state $\si[(n)]$, i.e.,
\begin{equation}
\si[(n)\perp]\propto\ha^{(n)}\si[(n)]-\si[(n)]\csi[(n)]\ha^{(n)}\si[(n)],
\end{equation}
which turns out to be
\begin{equation}
\si[(n)\perp]=\frac{1}{\sqrt{2}}(|a_{\max}\rangle^{\otimes n}-\et|a_{\min}\rangle^{\otimes n}).
\end{equation}
Therefore, the optimal postselected state is
\begin{equation}
\begin{aligned}\sff[(n)] & =\bigg(\sqrt{\frac{P_{s}}{2}}+\sqrt{\frac{1-P_{s}}{2}}\bigg)|a_{\max}\rangle^{\otimes n}\\
 & +\et\bigg(\sqrt{\frac{P_{s}}{2}}-\sqrt{\frac{1-P_{s}}{2}}\bigg)|a_{\min}\rangle^{\otimes n}.
\end{aligned}
\label{eq:30}
\end{equation}

The postselected state $\sff[(n)]$ (either (\ref{eq:16}) or (\ref{eq:30}))
is also an entangled state. Generally speaking, postselecting $n$
systems in an entangled state is very nontrivial. In Sec. \ref{sec:Qubit-example},
we show how to achieve this kind of postselection by simple quantum
circuits for qubits. That method can be generalized to higher dimensional
systems.

\section{Fisher information of weak measurement\label{sec:Fisher-information-of}}

Precision is one of the most important benchmarks for the performance
of a measurement. Since weak measurement can amplify small parameters
by postselecting the system in addition to measuring the pointer,
it has long been speculated that the postselected weak measurements
can increase the precision of measuring small parameters. However,
since failed postselection events comprise a large fraction of total
events, the precision can also be significantly reduced by low postselection
efficiency. It is possible that the increase of the precision by weak
value amplification may be canceled by low efficiency. Therefore,
whether postselection can really increase the precision of weak measurement
is controversial, and it is important to make clear how well weak
measurement with postselection can do in the metrology of small parameter
estimation. This has become a hot topic of recent research.

In this section, we study this problem in detail. We compute the Fisher
information, a widely used metric for the precision of parameter estimation,
for general weak measurements with postselection of the system, and
use the results from the last section to obtain the maximum Fisher
information for a given weak value or a fixed postselection probability.
With the assistance of entanglement among the systems, the Fisher
information may be increased approximately to the Heisenberg limit,
which is the upper bound for quantum Fisher information, and the loss
of Fisher information in the failed postselection events can be reduced
to the order of the small parameter, which is negligible in the regime
of the weak value approximation. So the performance of entanglement-assisted
weak value amplification essentially achieves the optimum for quantum
metrology.

\subsection{Background of Fisher information}

We first introduce the Fisher information. The precision of estimating
a parameter is usually quantified by the variance of the estimate.
But it is often not easy to compute the exact variance of an estimate,
since the estimate itself often does not have an analytical solution
for many estimation strategies. Fortunately, the Cram\'er-Rao relation
\cite{Cramer1946} gives a lower bound for the variance of an unbiased
estimate, quantified by the Fisher information.

For a probability distribution $p_{g}(x)$ dependent on a parameter
$g$, the minimum statistical error of estimating $g$ from $p_{g}(x)$
satisfies
\begin{equation}
\langle\delta g^{2}\rangle\geq\frac{1}{n\ig}+\langle\delta g\rangle^{2},\label{eq:31}
\end{equation}
where $n$ is the number of sample data, and $I_{g}$ is the Fisher
information defined as
\begin{equation}
\ig=\int_{X}\frac{(\partial_{g}p_{g}(x))^{2}}{p_{g}(x)}\d x,\label{eq:fisher info}
\end{equation}
where $X$ is the region that $x$ belongs to. If the estimation strategy
is unbiased, the estimate bias $\langle\delta g\rangle$ will vanish,
and the statistical error of the estimate is lower bounded by the
reciprocal of the Fisher information. It can be proved that the lower
bound (\ref{eq:31}) can be saturated in the limit $n\rightarrow\infty$
when the estimation uses the maximum likelihood strategy.

For a quantum state $|\Phi_{g}\rangle$ dependent on a parameter $g$,
the method to estimate $g$ is to measure many copies of $|\Phi_{g}\rangle$,
and estimate $g$ from the distribution of measurement results. Suppose
the measurement is described by a POVM $\{\he_{1},\cdots,\he_{r}\}$,
where
\begin{equation}
\he_{i}\geq0,\,\mathrm{and}\,\sum_{i}\he_{i}=I.
\end{equation}
Then the probability of obtaining the $i$-th result is
\begin{equation}
p(i)=\avg{\he_{i}}.\label{eq:13}
\end{equation}
Obviously, the probability of measurement results $p(i)$ is dependent
on $g$, so from the distribution $p(i)$ one can estimate the parameter
$g$. And the Fisher information of estimation can be obtained by
plugging (\ref{eq:13}) into (\ref{eq:fisher info}).

Since there are many different choices of measurement on $\ket{\fg}$,
which lead to different Fisher informations of estimating $g$, there
exists a maximum of the Fisher information over all choices of measurement.
This maximum Fisher information is called the \emph{quantum Fisher
information} \cite{Braunstein1994,Braunstein1996}, and is found to
be
\begin{equation}
\iq=4\big(\bk{\pg}{\pg}-|\bk{\fg}{\pg}|^{2}\big).\label{eq:2-1}
\end{equation}
In (\ref{eq:2-1}), the dependence on the choice of measurement has
vanished, and the quantum Fisher information $\ig[(Q)]$ is determined
solely by the state $|\Phi_{g}\rangle$. 

A common task in quantum metrology is to estimate some multiplicative
parameter $g$ of a Hamiltonian in the form $gH$. In this case, one
usually prepares a quantum system in some initial state $|\Phi\rangle$
and let it evolve under the Hamiltonian $gH$ for some time $t$.
The final state of the system is $\exp(-itg\hat{H})|\Phi\rangle$.
Then one can do a measurement on the system, and when the measurement
is optimized, the quantum Fisher information is \cite{Giovannetti2006}
\begin{equation}
\iq=4\text{Var}(\hat{H})_{|\Phi\rangle},\label{eq:1-1-1}
\end{equation}
which is entirely determined by the variance of the Hamiltonian $\hat{H}$
in the initial state $|\Phi\rangle$.

\subsection{General result for the Fisher information of weak measurements}

With the above background knowledge of quantum Fisher information,
we go on to study the precision of postselected weak measurements
in this subsection. Our central focus is to investigate whether the
competition between the amplification by the weak value and the reduction
by the low postselection probability leads to a gain or a loss of
the Fisher information, and how much the gain or the loss is. To achieve
this aim, we compare the quantum Fisher information of estimating
$g$ with and without postselection of the system.

In a weak measurement with (\ref{eq:12}) as the interaction Hamiltonian
and $\si,\,\di$ as the respective initial states of the system and
pointer, the whole system evolves to the joint state $\exp(-\i g\ha\otimes\hf)\si\otimes\di$
after the weak interaction. If there is no postselection of the system
after the interaction, then according to Eq. (\ref{eq:1-1-1}), the
quantum Fisher information is 
\begin{equation}
\iq=4\Big[\langle\ha^{2}\rangle_{\si}\langle\hf^{2}\rangle_{\di}-\big(\langle\ha\rangle_{\si}\langle\hf\rangle_{\di}\big){}^{2}\Big].\label{eq:maxinf}
\end{equation}

Now suppose we perform a projective measurement on the system in order
to make a postselection. This measurement will produce $d$ independent
outcomes corresponding to some orthonormal basis $\{\sff[k]\}_{k=1}^{d}$,
where $d$ is the dimension of the system. In the linear response
regime with $g\ll1$, each of these outcomes corresponds to a postselection
of the system, and collapses the pointer to the state
\begin{equation}
\df[k]\approx\big(\hat{I}-\i gA_{w}^{(k)}\hf\big)\di,
\end{equation}
with success probability $\ps[k]\approx|\langle\Psi_{f}^{(k)}|\Psi_{i}\rangle|^{2}$,
and
\begin{equation}
A_{w}^{(k)}=\frac{\langle\Psi_{f}^{k}|\hat{A}|\Psi_{i}\rangle}{\langle\Psi_{f}^{k}|\Psi_{i}\rangle}.
\end{equation}
Then according to (\ref{eq:2-1}), the Fisher information in each
of the collapsed states $\df[k]$ after postselecting the system to
$\sff[k]$ is
\begin{align}
\ig[(k)] & \approx4\,\ps[k]|A_{w}^{(k)}|^{2}\Big[\text{Var}(\hf)_{\di}\nonumber \\
 & -\langle\hf^{2}\rangle_{\di}\big(2g\text{Im}A_{w}^{(k)}\langle\hf\rangle_{\di}+g^{2}|A_{w}^{(k)}|^{2}\langle\hf^{2}\rangle_{\di}\big)\Big].\label{eq:52}
\end{align}

It is important to observe that if we add the information from all
$d$ postselections, we obtain 
\begin{align}
\sum_{k=1}^{d}\ig[(k)] & \approx4\langle\ha^{2}\rangle_{\si}\text{Var}(\hf)_{\di}-O(g),\label{eq:1-2}
\end{align}
where we have used
\begin{equation}
\sum_{k}\ps[k]|A_{w}^{(k)}|^{2}=\langle\ha^{2}\rangle_{\si}.
\end{equation}
With the condition $\langle\hf\rangle_{\di}=0$, $\var(\hf)_{\di}=\langle\hf^{2}\rangle_{\di}$,
then (\ref{eq:1-2}) saturates the maximum in (\ref{eq:maxinf}) up
to a small corrections of order $g$, which indicates that the measurement
on the system does not lose information by itself, but rather redistributes
and concentrates the information about $g$ in the final pointer states
$\{\df[k]\}_{k=1}^{d}$.

In a postselected weak measurement, if we postselect the system in
the state $\sff[k]$, we discard all events in which the system does
not collapse to $\sff[k]$. So for such a postselected weak measurement,
the Fisher information where the pointer is initially in an unbiased
state (i.e., $\langle\hf\rangle_{\di}=0$) is
\begin{equation}
\ig[(k)]\approx4\ps[k]|A_{w}^{(k)}|^{2}\langle\hf^{2}\rangle_{\di}\big(1-g^{2}|A_{w}^{(k)}|^{2}\langle\hf^{2}\rangle_{\di}\big).\label{eq:32}
\end{equation}

It is worth mentioning that Ref. \cite{Tanaka2013} derived results
about the total Fisher information of postselected weak measurement
similar to Eqs. (\ref{eq:52}) and (\ref{eq:32}), though in a slightly
different notation, and Ref. \cite{Ferrie2014a} obtained a result
similar to (\ref{eq:1-2}), including the classical Fisher information
of the postselected result distribution.

\subsection{Maximum Fisher information of postselected weak measurements}

Having obtained the general result for quantum Fisher information
of a postselected weak measurement in Eq. (\ref{eq:32}), we now go
on to consider maximizing the quantum Fisher information from a postselected
weak measurement, and investigate whether, and how much, Fisher information
is lost by discarding the failed postselection events. This is currently
the subject of hot debate.

From Eq. (\ref{eq:32}) we see that the dependence of the Fisher information
$\ig[(k)]$ on the initial and postselected states of the system is
determined by the postselection probability $\ps[k]$ and the weak
value $A_{w}^{(k)}$. To maximize the Fisher information $\ig[(k)]$,
we use the results on the maximum postselection probability given
the weak value, or maximum weak value given the postselection probability,
that were obtained in Sec. \ref{sec:Optimization}.

From Sec. \ref{sec:Optimization}, we know that when the weak value
is sufficiently large, i.e., $|A_{w}|\gg\max\lambda(\ha)$, then the
maximum $P_{s}$ and the maximum $|A_{w}|$ can be approximated as
\begin{equation}
\begin{aligned}\max P_{s} & \approx\frac{\text{Var}(\ha)_{\si}}{|A_{w}|^{2}}, & \text{with}\,A_{w}\,\text{fixed},\\
\max|A_{w}| & \approx\sqrt{\frac{\text{Var}(\ha)_{\si}}{P_{s}}}, & \text{with}\,P_{s}\,\text{fixed}.
\end{aligned}
\label{eq:33-1}
\end{equation}

Now, we can plug either equation of (\ref{eq:33-1}) into (\ref{eq:32}),
and obtain
\begin{equation}
\max\ig[(k)]\approx4\text{Var}(\ha)_{\si}\langle\hf^{2}\rangle_{\di}\big(1-g^{2}|A_{w}^{(k)}|^{2}\langle\hf^{2}\rangle_{\di}\big).\label{eq:34}
\end{equation}
Comparing this maximum Fisher information for weak measurement with
postselection to that without postselection in (\ref{eq:maxinf}),
it follows that
\begin{equation}
\max\ig[(k)]\approx\ig[(Q)]\frac{\text{Var}(\hat{A})_{|\Psi_{i}\rangle}}{\langle\hat{A}^{2}\rangle_{|\Psi_{i}\rangle}}\big(1-|gA_{w}^{(k)}|^{2}\langle\hf^{2}\rangle_{\di}\big),\label{eq:kinfopt}
\end{equation}
where $\ig[(Q)]$ is the global quantum Fisher information with an
unbiased pointer ($\langle\hf\rangle_{\di}=0$). And it is almost
equal to $\iq$ (\ref{eq:maxinf}) if the system is initially unbiased
($\langle\ha\rangle_{\si}=0$).

This implies that postselection redistributes the quantum Fisher information
between the system and the pointer, and with the optimal choice of
pre- and postselection of the system, it can concentrate nearly all
the Fisher information into a single (but very improbable) pointer
state. The remaining very small amount of information is distributed
among the failed postselection events, and could be retrieved in principle
by measuring the respective collapsed pointer states. The pointer
state corresponding to successful postselection of the system suffers
an overall reduction factor of $\text{Var}(\hat{A})/\langle\hat{A}^{2}\rangle$
which is $1$ for unbiased system states, as well as a tiny loss $|gA_{w}^{(1)}|^{2}\langle\hf^{2}\rangle_{\di}$.
However, most weak value amplification experiments operate in the
linear response regime $g|A_{w}^{(1)}|\langle\hf^{2}\rangle_{\di}^{\frac{1}{2}}\ll1$,
so this remaining loss is negligible. Moreover, the overall reduction
can be further compensated by extracting information from the postselection
probability distribution \cite{Zhang2015}.

This is quite a surprising result because it implies that one can
approximately saturate the global optimal bound of Fisher information
(\ref{eq:maxinf}) by measuring only the very rare postselected pointer
state while the remaining much more probable outcomes are discarded,
although the optimal bound (\ref{eq:maxinf}) cannot be exactly reached
\cite{Tanaka2013,Ferrie2014a,Zhang2015}. The unlikely postselections
can also offer an advantage in practice: in measuring small signals
in the face of experimental imperfections, it can be easier to see
rare large events than frequent small ones. This property of postselected
weak measurement makes it a broadly useful technique for estimating
small parameters within the linear response regime \cite{Jordan2014}.

\subsection{Saturation of Heisenberg limit\label{sub:Saturation-of-Heisenberg}}

A well-known upper bound on the quantum Fisher information is the
Heisenberg limit, which sets the ultimate upper bound for the sensitivity
of quantum parameter estimation, and demonstrates that the parameter
estimation precision by quantum measurements can scale as $n^{-1}$
(or equivalently, quantum Fisher information can scale quadratically
with $n$), if $n$ quantum systems are coupled. This is higher than
the standard quantum limit (SQL) (or the classical limit) $n^{-\frac{1}{2}}$,
by order $\sqrt{n}$.

An interesting question in weak value amplification is whether weak
measurement with postselection can achieve Heisenberg-limited scaling
in precision. As we showed in the last subsection, the quantum Fisher
information of postselected weak measurement can be approximately
promoted to the global maximum quantum Fisher information $\iq$ by
optimized pre- and postselections of the system. The global quantum
Fisher information can generally achieve the Heisenberg limit using
entangled systems (and pointers), so we would expect that the Fisher
information of postselected weak measurement can also reach the Heisenberg
limit. 

We can straightforwardly verify this idea by exploiting the results
that were obtained in the previous sections. Similar to standard quantum
metrology, when the $n$ systems are uncorrelated the Fisher information
of a postselected weak measurement scales like the standard quantum
limit, i.e., $n$, while with entanglement among the systems, the
Fisher information can be boosted to scale like the Heisenberg limit
$n^{2}$.

In Eq. (\ref{eq:34}), it was shown that the maximum Fisher information
of a postselected weak measurement is proportional to the variance
of the system observable $\ha$ in the initial state of the system.
Therefore, the scaling of the variance determines the scaling of the
Fisher information.

Suppose we have $n$ systems. From Sec. \ref{sub:Uncorrelated-systems}
and \ref{sub:Entangled-systems}, we know that the variance of the
total observable $\ha^{(n)}$ scales linearly with $n$ when the $n$
systems are initially uncorrelated, and scales quadratically with
$n$ when the $n$ systems are initially entangled. Therefore, we
can immediately conclude that the maximum Fisher information of a
postselected weak measurement with $n$ systems can indeed scale like
the Heisenberg limit $n^{2}$, and the necessary ingredient to reach
this limit is entanglement between the systems initially.

The achievability of the Heisenberg limit by using entanglement can
also be understood in another way. It is known that the total Fisher
information of estimating a parameter from a sample of data is proportional
to the size of the sample. Since failed postselection events are discarded
in postselected weak measurements, the total Fisher information is
proportional to the postselection probability. We know from Sec. \ref{sec:Optimization}
that the postselection probability scales linearly with $n$ when
the $n$ systems are initially uncorrelated, and scales quadratically
with $n$ when the $n$ systems are initially entangled. Therefore,
the total Fisher information can scale as $n^{2}$ if the initial
state of the $n$ systems is entangled.

The above simple result again verifies the previous conclusion that
with optimized pre- and postselections of the system, the Fisher information
of a postselected weak measurement can approximately reach the global
optimal bound, and the loss of Fisher information by discarding the
failed postselection events can be negligible.

\section{Qubit example\label{sec:Qubit-example}}

To illustrate the new protocol of entanglement-assisted weak measurement,
we give an example with qubits in this section. For clarity, we will
focus on the protocol for increasing the postselection probability
with a fixed weak value from now on. The case of increasing the weak
value with a fixed postselection probability can be derived straightforwardly.

\subsection{Protocol}

Suppose we use $n$ qubits as systems and let them couple to a common
pointer qubit. The interaction Hamiltonian between each system qubit
and the pointer qubit is
\begin{equation}
H_{{\rm int}}=\varphi\sz\otimes\sx,\,\varphi\ll1.\label{eq:18}
\end{equation}

\begin{figure}
\begin{centering}
\includegraphics[scale=1.1]{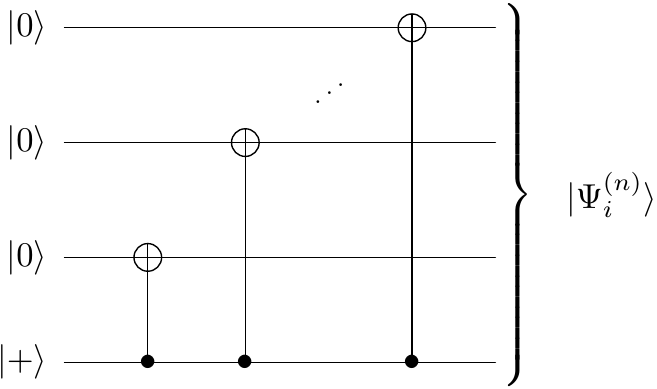}\protect\caption{Quantum circuit to initialize the $n$ system qubits. The qubits are
prepared in the maximally entangled state $|\Psi_{i}\rangle=(|0\rangle^{\otimes n}+|1\rangle^{\otimes n})/\sqrt{2}$
by a sequence of CNOT gates. }

\par\end{centering}

\label{fig:preparation} 
\end{figure}

In the entanglement-assisted weak measurement scheme, the initialization
step prepares the $n$ qubits in a maximally entangled state:
\begin{equation}
\si[(n)]=\frac{1}{\sqrt{2}}(\kon+\kln).\label{eq:15}
\end{equation}
This can be achieved by inputting $n-1$ qubits in the state $\ko$
and one qubit in the state $\kl$ into the circuit in Fig. \ref{fig:preparation}.
A sequence of $n-1$ CNOT gates entangles the $n$ qubits, and produces
the desired maximally entangled state (\ref{eq:15}).

Next, the $n$ qubits are subject to the weak interaction with the
common pointer qubit (\ref{eq:18}). To simulate this interaction,
note that the unitary interaction under $H_{{\rm int}}$ can be written
as
\begin{equation}
\begin{aligned}\hu & =\exp(-\i\varphi\sz\otimes\sx)\\
 & =|0\rangle\langle0|\otimes\exp(-\i\varphi\sx)+|1\rangle\langle1|\otimes\exp(\i\varphi\sx)\\
 & =(|0\rangle\langle0|\otimes I+|1\rangle\langle1|\otimes\exp(2\i\varphi\sx))(I\otimes\exp(-\i\varphi\sx)),
\end{aligned}
\end{equation}
so $\hu$ can be realized with a control-$\hrx(-4\varphi)$ rotation
followed by a $\hrx(2\varphi)$ gate on the pointer qubit, where $\hrx(\theta)$
\cite{Nielsen2000} is defined as
\begin{equation}
\hrx(\theta)=\exp\Big(-\i\frac{\theta}{2}\sx\Big).
\end{equation}
Since the control-$\hrx(-4\varphi)$ rotation and the $\hrx(2\varphi)$
rotation commute, all the $\hrx(2\varphi)$ gates can be delayed until
after the last control-$\hrx(-4\varphi)$ rotation. So the total interaction
$U^{\otimes n}$ is equivalent to by $n$ control-$\hrx(-4\varphi)$
rotations between the system qubits and the pointer qubit followed
by an $\hrx(2n\varphi)$ gate on the pointer.

\begin{figure}
\begin{centering}
\includegraphics[scale=0.9]{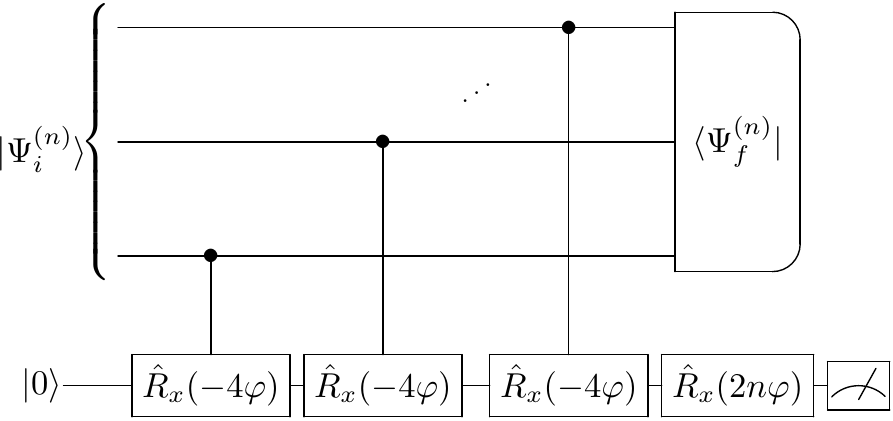}\protect\caption{Quantum circuit equivalent to the weak interaction $H_{{\rm int}}=\varphi\protect\sz\otimes\protect\sx,\,\varphi\ll1,$
between $n$ system qubits and one common pointer qubit.}

\par\end{centering}

\label{fig:circuitentangled} 
\end{figure}

Therefore, the weak interaction in the weak measurement using $n$
system qubits and one common pointer qubit can be implemented by the
circuit in Fig. \ref{fig:circuitentangled}.

After the weak interactions, the $n$ system qubits are postselected
to the state $\sff[(n)]$ (\ref{eq:16}), which in this case turns
out to be
\begin{equation}
\sff[(n)]\propto(n+A_{w}^{*})|0\rangle^{\otimes n}+(n-A_{w}^{*})|1\rangle^{\otimes n}.\label{eq:17}
\end{equation}
Such a postselection realizes a given weak value $A_{w}$. 

Postselection of $n$ system qubits in an entangled state $\sff[(n)]$
(\ref{eq:17}) is usually not easy, but it can be realized as the
inverse procedure for a preparation. Note that $\sff[(n)]$ can be
written as
\begin{equation}
\sff[(n)]=\hv\kon,\label{eq:41}
\end{equation}
where $\hv$ is a unitary transformation that turns $\kon$ to $\sff[(n)]$.
$\hv$ can be realized by CNOT gates and single qubit gates that are
similar to the preparation of $\si[(n)]$. 

There are many different choices of $\hv$ to realize (\ref{eq:41}).
One convenient choice among them is to transform the subspace spanned
by $\{\kon,\kln\}$ to itself, i.e., the subspace spanned by $\{\kon,\kln\}$
is invariant under $\hv$. The advantage of such a choice is that
there are only two possible postselected states of the $n$ system
qubits, $\kon$ and $\kln$, and these two states have the largest
Hamming distance, which will be helpful to fighting against readout
errors that will be discussed in Sec. \ref{sec:Readout-errors}. 

\begin{figure}
\begin{centering}
\includegraphics[scale=0.9]{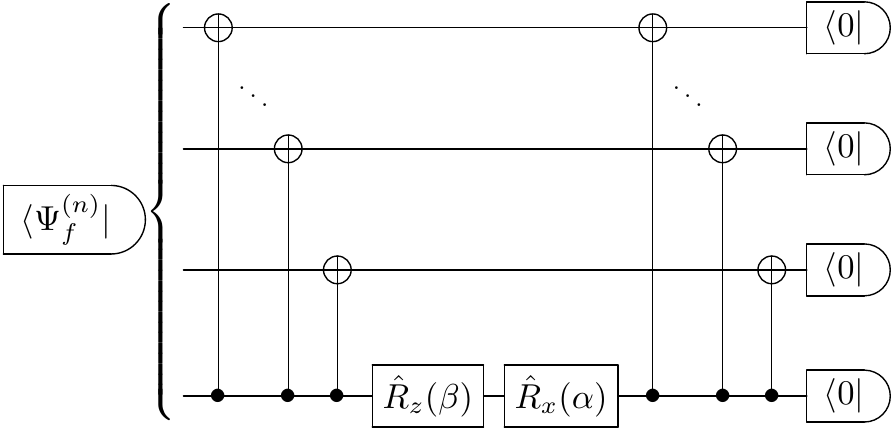}\protect\caption{Quantum circuit to postselect the $n$ system qubits in the entangled
state $\protect\sff$ (\ref{eq:17}), which is in essence the reverse
process of preparing $\protect\sff$ from $|0\rangle^{\otimes n}$.}

\par\end{centering}

\label{fig:postselection} 
\end{figure}

Such a $\hv$ can be written as
\begin{equation}
\hv:\;\begin{cases}
\kon\longrightarrow\sff[(n)],\\
\kln\longrightarrow\sff[(n)\perp],
\end{cases}\label{eq:20}
\end{equation}
where
\begin{equation}
\sff[(n)\perp]\propto(n+A_{w})\kln-(n-A_{w})\kon.\label{eq:1}
\end{equation}
Then, the postselection of the $n$ system qubits to $\sff[(n)]$
can be decomposed into the reverse unitary transformation $\hv^{\dagger}$
followed by a postselection in the state $\kon$. It can be implemented
by the circuit in Fig. \ref{fig:postselection}, where
\begin{equation}
\begin{aligned}\alpha & =-2\arctan\sqrt{\frac{n^{2}+|A_{w}|^{2}-2n\re A_{w}}{n^{2}+|A_{w}|^{2}+2n\re A_{w}}},\\
\beta & =-\frac{\pi}{2}-\arg\frac{n-A_{w}^{*}}{n+A_{w}^{*}}.
\end{aligned}
\end{equation}

In the following, we compute the composite state of the $n$ system
qubits and the pointer qubit before the postselection, and the probability
of successful postselection. These results will be useful in the remainder
of the paper.

Suppose the initial state of the $n$ system qubits is $\si=(\kon+\kln)/\sqrt{2}$.
Then after the weak interaction, the joint state of the system qubits
and the pointer qubit is
\begin{equation}
\begin{aligned}|\Phi\rangle & =\hv^{\dagger}\exp(-\i\varphi(\hat{\sigma}_{1z}+\cdots+\hat{\sigma}_{nz})\otimes\sx)\si\di\\
 & \propto(\kon\csf[(n)]+\kln\csf[(n)\perp])\\
 & (\kon\e^{-\i n\varphi\sx}+\kln\e^{\i n\varphi\sx})\di.
\end{aligned}
\end{equation}
According to (\ref{eq:17}) and (\ref{eq:1}), $|\Phi\rangle$ can
be simplified to
\begin{equation}
\begin{aligned}|\Phi\rangle & \propto\kon((n+A_{w})\e^{-\i n\varphi\sx}+(n-A_{w})\e^{\i n\varphi\sx})\di\\
 & +\kln((-n+A_{w}^{*})\e^{-\i n\varphi\sx}+(n+A_{w}^{*})\e^{\i n\varphi\sx})\di.
\end{aligned}
\label{eq:21}
\end{equation}
So, when the $n$ system qubits are postselected in $\kon$, the pointer
state collapses to
\begin{equation}
\df[,0]=(n\cos n\varphi-\i A_{w}\sin n\varphi\sx)\di,\label{eq:23}
\end{equation}
and when the $n$ system qubits are postselected in $\kln$, the pointer
qubit collapses to
\begin{equation}
\df[,1]=(A_{w}^{*}\cos n\varphi+\i n\sin n\varphi\sx)\di.\label{eq:24}
\end{equation}

The postselection probabilities for $\kon$ and $\kln$ can be worked
out from (\ref{eq:21}) respectively:
\begin{equation}
\begin{aligned}\pon & =\frac{\eta_{0}}{\eta_{0}+\eta_{1}},\\
\pln & =\frac{\eta_{1}}{\eta_{0}+\eta_{1}},
\end{aligned}
\label{eq:2}
\end{equation}
where the superscripts $(n)$ denote there are $n$ entangled qubits,
and
\begin{equation}
\begin{aligned}\eon & =n^{2}\cos^{2}n\varphi+|A_{w}|^{2}\sin^{2}n\varphi+n\im A_{w}\sin2n\varphi\langle\sx\rangle_{D},\\
\eln & =|A_{w}|^{2}\cos^{2}n\varphi+n^{2}\sin^{2}n\varphi-n\im A_{w}\sin2n\varphi\langle\sx\rangle_{D}.
\end{aligned}
\label{eq:3}
\end{equation}

\subsection{Fisher information and Heisenberg limit}

In this subsection, we calculate the Fisher information for the qubit
example. We will see that with entanglement between the system qubits,
the Fisher information of the pointer qubit can indeed reach the Heisenberg
limit and saturate the Cram\'er-Rao bound.

Based on the the final pointer state and the postselection probability
obtained in the last subsection, we can compute the quantum Fisher
information of an entanglement-assisted postselected weak measurement.
To simplify the computation, we assume that $n|\varphi|\ll1,\,|A_{w}\varphi|\ll1$,
and $n^{2}|\varphi|\ll|A_{w}|$ which usually hold in the weak value
approximation. Then $\df[0]$ and $\df[1]$ can be simplified to
\begin{equation}
\begin{aligned}\df[0] & \approx(I-\i A_{w}\varphi\sx)\di,\\
\df[1] & \approx\Big(I+\i n^{2}\frac{\varphi}{A_{w}^{*}}\sx\Big)\di,
\end{aligned}
\label{eq:2-2}
\end{equation}
and
\begin{equation}
\pon\approx\frac{n^{2}}{|A_{w}|^{2}+n^{2}},\,\pln\approx\frac{|A_{w}|^{2}}{|A_{w}|^{2}+n^{2}}.\label{eq:8-1}
\end{equation}

According to (\ref{eq:2-2}), $|\partial_{\varphi}D_{f0}\rangle\approx-\i A_{w}\sx\di,$
so from the definition of quantum Fisher information (\ref{eq:2-1}),
and taking the postselection probability into consideration, the quantum
Fisher information of the final pointer state when the $n$ qubits
are all postselected to $\ko$ becomes
\begin{equation}
\ig[(\kon)]=\frac{4n^{2}|A_{w}|^{2}(1-|A_{w}|^{2}\varphi^{2})}{n^{2}+|A_{w}|^{2}}.
\end{equation}
When $|A_{w}|\varphi\ll1$ and $|A_{w}|\gg n$,
\begin{equation}
\ig[(\kon)]\approx4n^{2}.
\end{equation}
This shows that the Heisenberg limit is approximately attained in
this case, and Fisher information is only lost to order $\varphi^{2}$.
Since $\varphi\ll1$ in the weak coupling regime, the loss of Fisher
information is negligible. Therefore, the Fisher information of a
postselected weak measurement can indeed approach the Heisenberg limit
with optimal pre- and postselections of the systems, and the final
state of the pointer qubit possesses almost all the Fisher information
of the phase $\varphi$. This verifies the result in Sec. \ref{sub:Saturation-of-Heisenberg}.

\section{Readout errors\label{sec:Readout-errors}}

In the previous sections, we introduced entanglement-assisted weak
measurement and studied its performance in metrology. We now turn
to issues that will arise in practical implementations.

Errors are inevitable in any practical application of a quantum protocol.
They can be caused by noise in the environment, or by technical imperfections.
Numerous ways have been invented to fight against errors, and systematic
theories, such as quantum error correction code and dynamical decoupling
\cite{Lidar2013}, have been developed to utilize them. 

In this section, we study a typical kind of error in postselected
weak measurements: readout errors. Readout error can result from noise
in the environment and technical imperfections in the experimental
devices. We focus on the readout errors that occur in the postselection
stage. This kind of error can significantly influence the weak measurement
protocol, by distorting the postselected results. As we shall see
later, even when the probability of a readout error is very low, the
disturbance to the postselection measurement can be dramatic. So correcting
this type of error is necessary in postselected weak measurements.

Readout errors influence postselection results in two main ways. First,
they may read some failed postselections as successful ones, which
can bring errors to the statistics of the postselection results. Second,
they may read some successful postselections as failed ones, which
will reduce the postselection efficiency.

Below, we start from the qubit example in Sec. \ref{sec:Qubit-example}
and analyze the effects of readout errors in postselection. We will
show that an initially entangled state of the $n$ system qubits can
dramatically increase the robustness of weak measurement, which demonstrates
another advantage of entanglement. We will also study the loss rate
of successful postselections, and its complementarity relation with
the error rate. In addition, the effect of readout errors on the Fisher
information will be considered. To fight against this type of error,
we introduce a majority vote scheme to reduce both the error rate
and the loss rate in the postselection results. This simple trick
can eliminate almost all loss of Fisher information caused by readout
errors.

\subsection{Error in postselection results\label{sub:Relative-error-rate}}

In this subsection, we analyze the first effect of readout errors
on a postselected weak measurement: that is, the relative error rate
of the successful postselection results. (We will omit the adjective
``relative'' when there is no ambiguity in the context.) The second
effect of readout errors, i.e., the loss of correct postselection
results, will be discussed in the next subsection.

Suppose the probability of mistaking $\ko$ for $\kl$ is $\qol$
and the probability of mistaking $|1\rangle$ for $|0\rangle$ is
$\qlo$. Then in an ordinary weak measurement, the probability of
obtaining a $\ko$ from reading a system qubit is
\begin{equation}
p(\ko)=\pol(1-\qol)+\pll\qlo.
\end{equation}
The component $\pll\qlo$ is the readout error which identifies $\kl$
as $\ko$, so the relative error rate in the postselection results
is
\begin{equation}
\peo=\frac{\pll\qlo}{\pol(1-\qol)+\pll\qlo}.\label{eq:26}
\end{equation}

Both $\qlo$ and $\qol$ are usually small. However, when $A_{w}\gg1$,
$\peo$ can become very large in some cases. To see this, use $\pll=1-\pol$
in (\ref{eq:26}), then (\ref{eq:26}) becomes
\begin{equation}
\peo=\frac{(1-\pol)\qlo}{\pol(1-\qol)+(1-\pol)\qlo}.\label{eq:36}
\end{equation}
The error rate $\peo$ can range from $0$ to $1$ when $p_{0}$ goes
from $1$ to $0$. 

In particular, if $\pol=\qol=\qlo$, then
\begin{equation}
\peo=\frac{1}{2};
\end{equation}
and if $\pol\ll\qlo$, then
\begin{equation}
\peo\rightarrow1.
\end{equation}
These imply that the postselected weak measurement is very sensitive
to readout errors, especially when the readout error probability is
comparable to the postselection probability.

In a postselected weak measurement, the postselection probability
is usually very small, so even a low probability of readout errors
may lead to a significant error in the postselection results. Therefore,
correcting readout errors or suppressing their influence is crucial
to practical applications of postselected weak measurement. 

Now let's consider the entanglement-assisted protocol of weak measurement,
and see whether the relative error rate in the postselection results
can be improved. 

In this protocol, the probability of correctly identifying a $\kon$
in the postselection is $(1-\qol)^{n}$, and the probability of mistaking
a $\kln$ for $\kon$ is $\qlo[n]$, so the total probability of reading
out $\kon$ from the $n$ system qubits after the weak coupling is
\begin{equation}
p(\kon)=\pon(1-\qol)^{n}+\pln\qlo[n].
\end{equation}
Similar to the above, the erroneous proportion of the postselection
results is $p_{1}\qlo[n]$, so the relative error probability in the
postselection results $|0\rangle^{\otimes n}$ is
\begin{equation}
\pe=\frac{\pln q_{1\rightarrow0}^{n}}{\pon(1-\qol)^{n}+\pln\qlo[n]}.\label{eq:4}
\end{equation}

\begin{figure}
\subfloat[\label{fig:pe-a}$\protect\qlo=\protect\qol=0.05$]{\protect\includegraphics[scale=0.47]{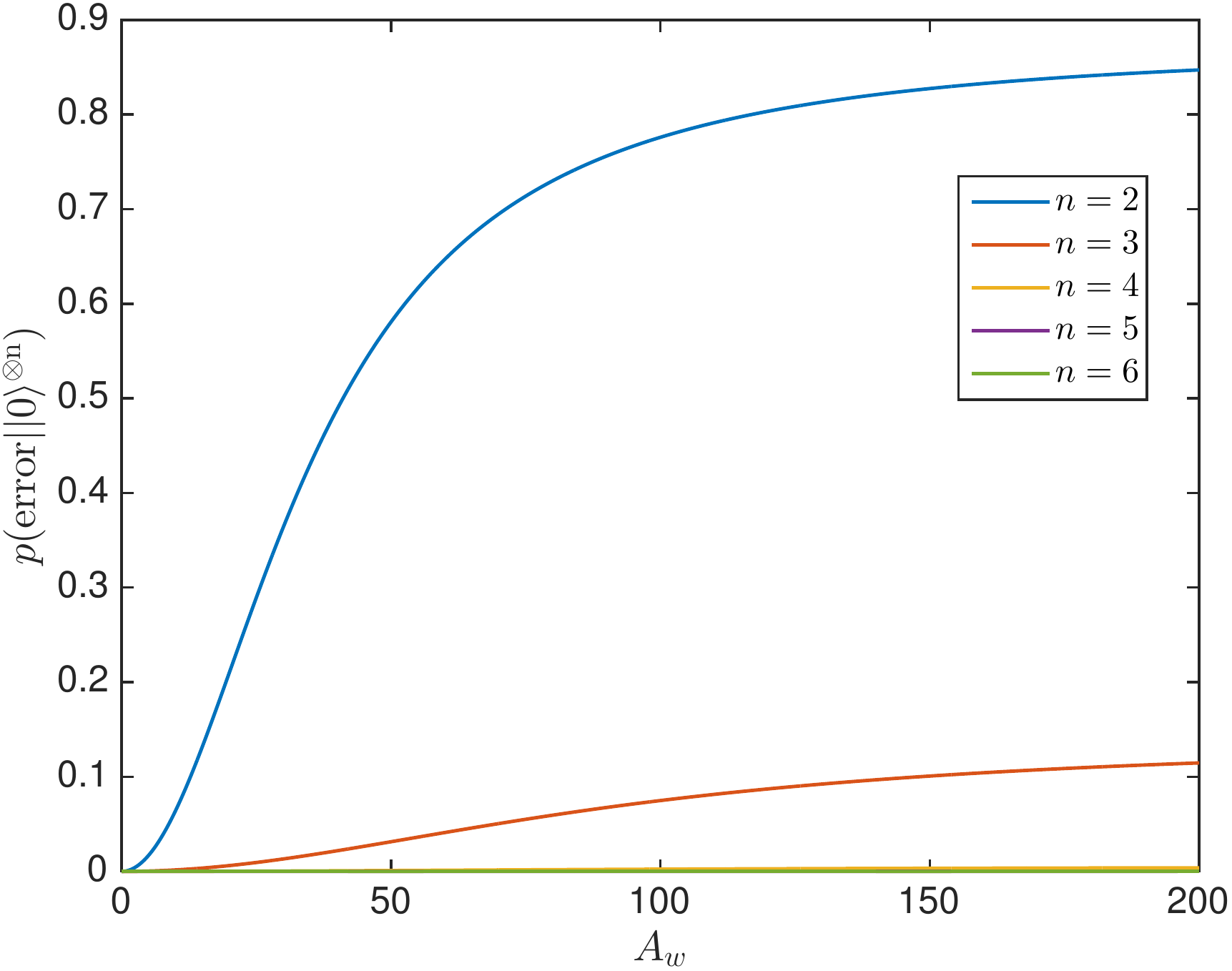}}\\\subfloat[\label{fig:pe-b}$\protect\qlo=\protect\qol=0.01$]{\protect\includegraphics[scale=0.47]{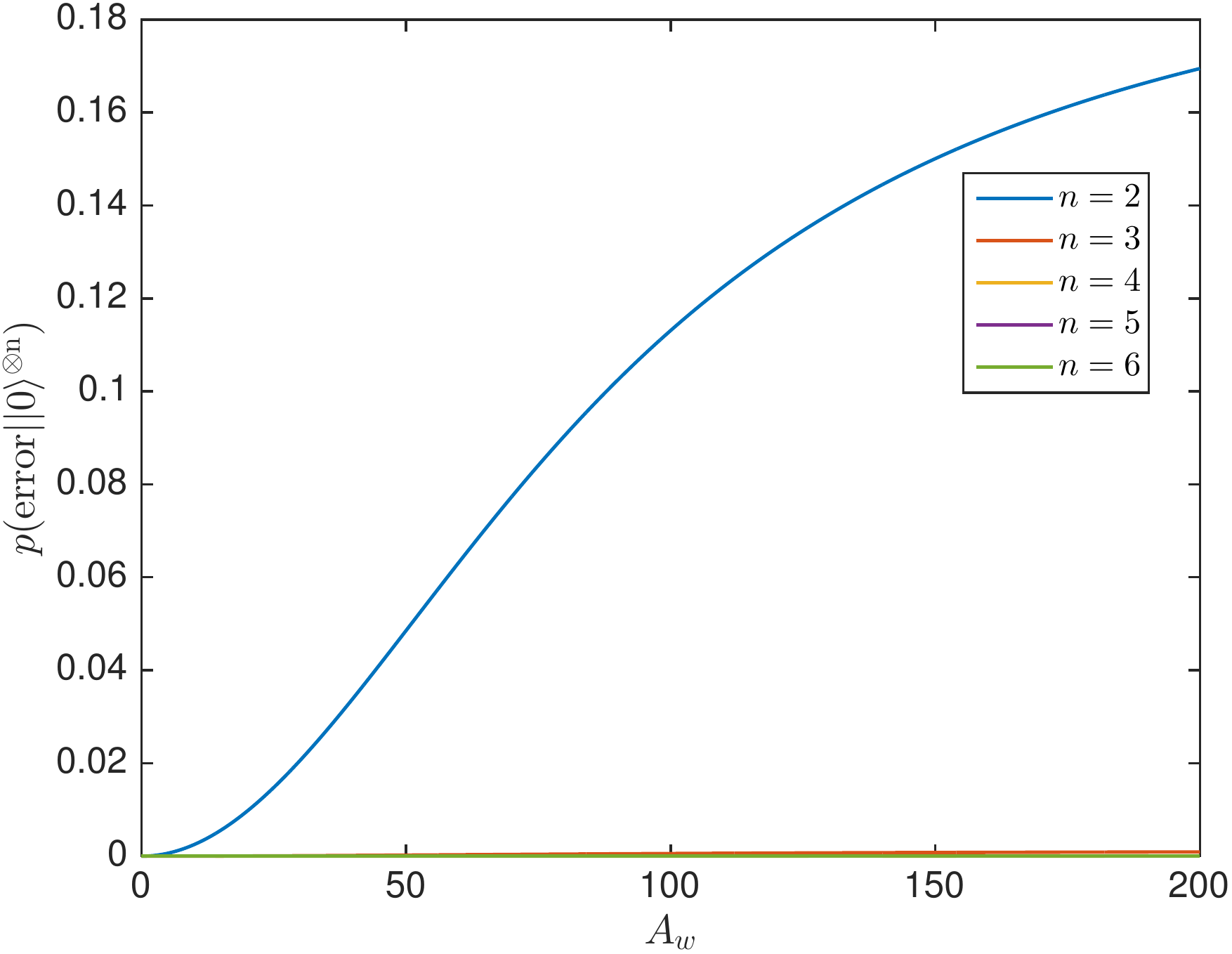}}\protect\caption{\label{fig:pe}(Color online) Plot of the relative error probability
$p(\mathrm{error}||0\rangle^{\otimes n})$ versus the weak value $A_{w}$
for different $n$. The lines from left to right in each figure are
for $n=2,\,3,\,4,\,5,\,6,$ respectively. (\ref{fig:pe-a}) When $\protect\qlo=\protect\qol=0.05$,
the $\protect\pe$ for $n=4,\,5,\,6$ are much lower than $n=2,\,3$,
and they almost overlap since they are very close to each other. (\ref{fig:pe-b})
Similarly, when $\protect\qlo=\protect\qol=0.01$, the $\protect\pe$
for $n=3,\,4,\,5,\,6$ are much lower than $n=2$, and they almost
overlap with each other.}
\end{figure}

In Fig. \ref{fig:pe}, we plot the relative error probability $\pe$
versus the weak value $A_{w}$ for different numbers $n$ of entangled
systems. We assume $\qlo$ and $\qol$ to be the same for simplicity.
As $n$ increases, the relative error $\pe$ dramatically decreases.

Note that $\pe$ can be rewritten as
\begin{equation}
\pe=\bigg(1+\frac{\pon}{\pln}\Big(\frac{1-\qol}{\qlo}\Big)^{n}\bigg)^{-1}.\label{eq:23-1}
\end{equation}
From (\ref{eq:2}), (\ref{eq:3}), when $n\varphi A_{w}\ll1$, we
have
\begin{equation}
\frac{\pon}{\pln}\approx\frac{n^{2}}{|A_{w}|^{2}}.
\end{equation}
Since $\qol,\qlo$ are usually very small, if $n$ further satisfies
that $\frac{1-\qol}{\qlo}\gg\sqrt[n]{\frac{|A_{w}|^{2}}{n^{2}}}$,
then Eq. (\ref{eq:23-1}) can be approximately simplified to
\begin{equation}
\pe\approx|A_{w}|^{2}n^{-2}\Big(\frac{1-\qol}{\qlo}\Big)^{-n}.\label{eq:25}
\end{equation}
This means that entanglement between the initial system qubits can
reduce the relative error rate caused by readout errors super-exponentially
with the number of entangled qubits! It implies how efficiently entanglement
can improve the robustness of postselected weak measurement against
the readout errors.

\begin{figure}
\subfloat[$\protect\qlo=\protect\qol=0.05$]{\protect\includegraphics[scale=0.47]{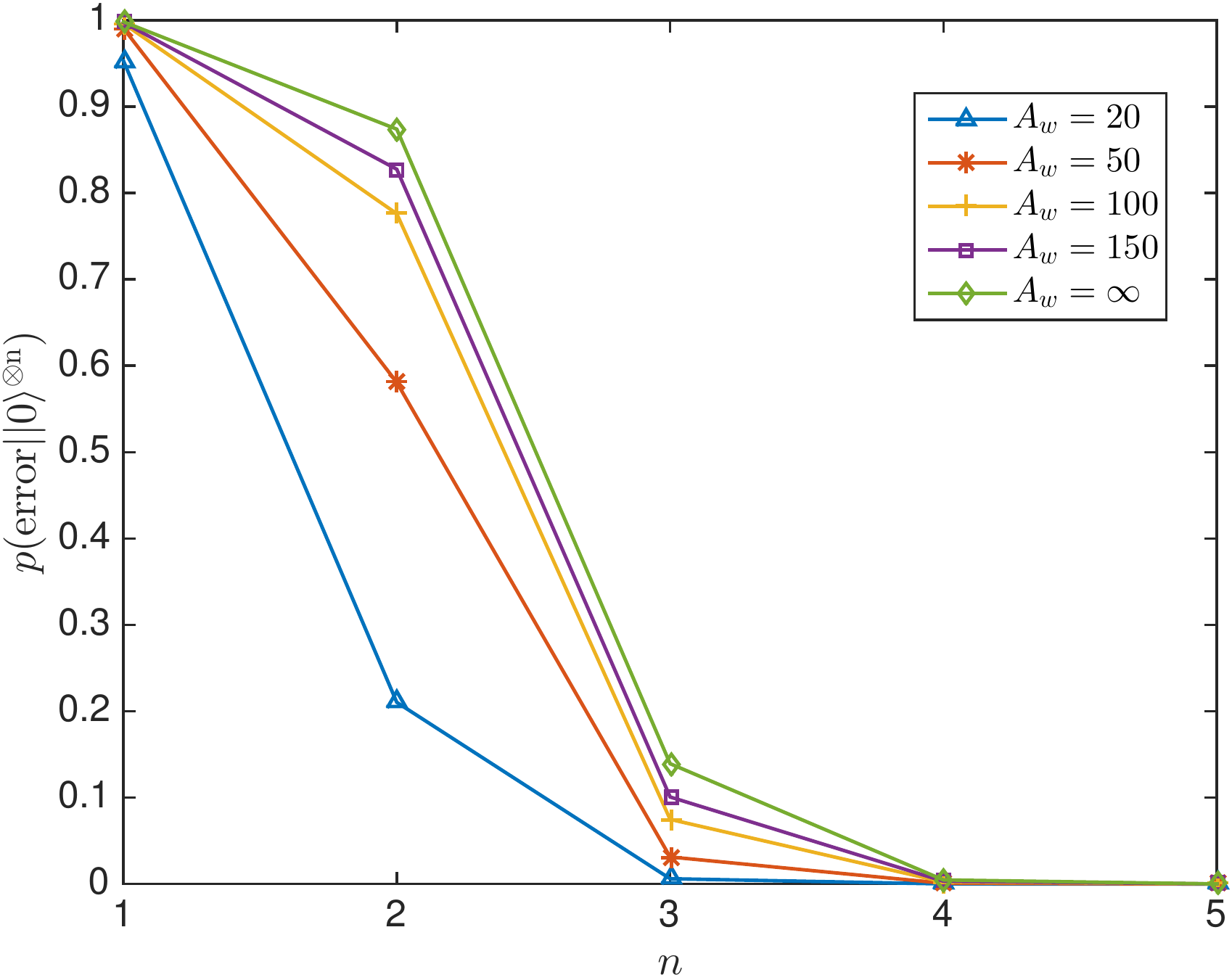}}\\\subfloat[$\protect\qlo=\protect\qol=0.01$]{\protect\includegraphics[scale=0.47]{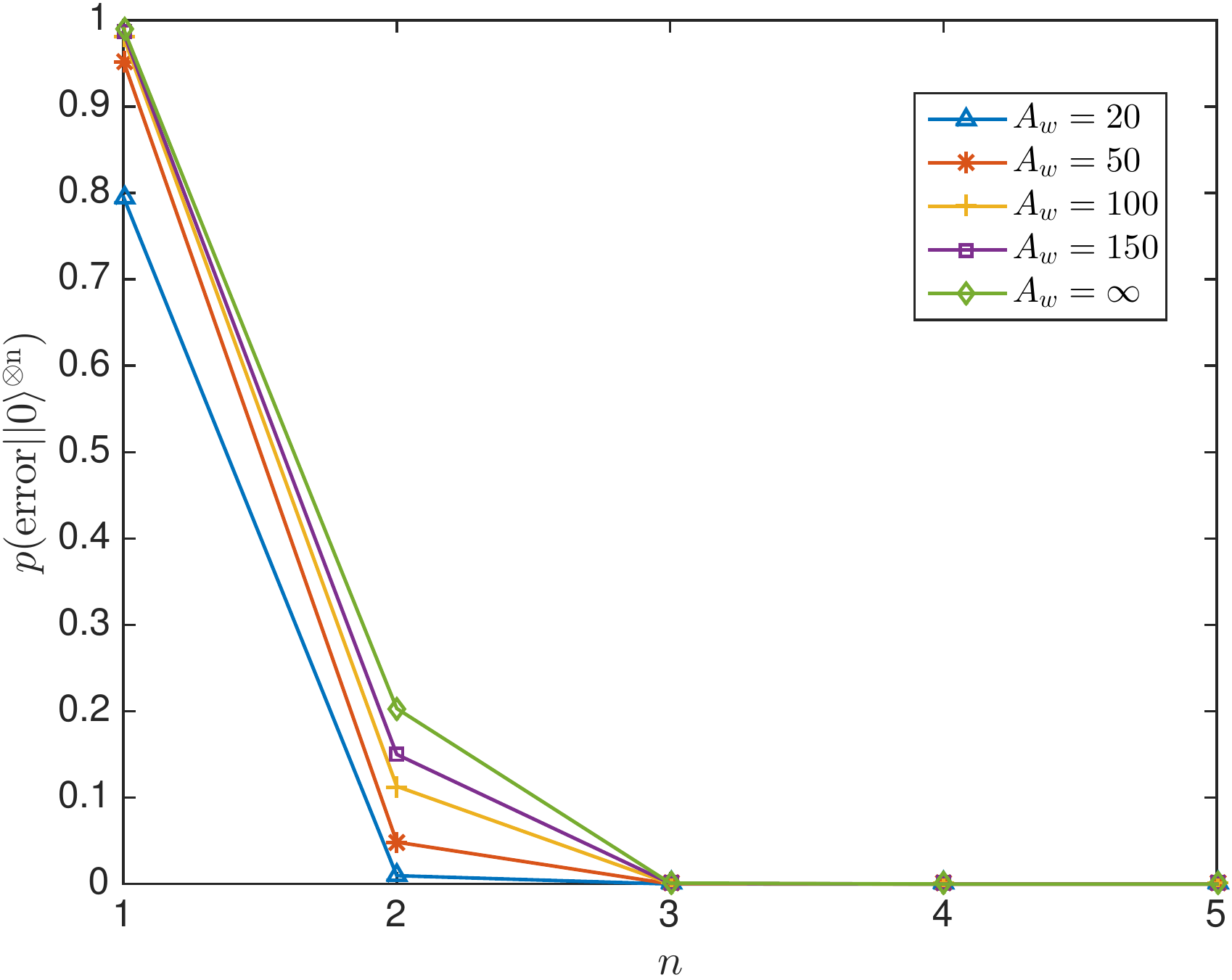}}\protect\caption{\label{fig:pen}(Color online) This figure plots the relative error
probability $p(\mathrm{error}||0\rangle^{\otimes n})$ versus the
number of entangled qubits $n$ for different weak values $A_{w}$.
It shows that the relative error rate $\protect\pe$ can decrease
very fast with $n$, which implies the advantage of entanglement in
suppressing the relative error rate.}
\end{figure}

To see this super-exponential decay of the relative error rate more
clearly, the relative error rate $\pe$ against $n$ is plotted in
Fig. \ref{fig:pen} for different weak values. In the figure, when
$n$ is small, $\pe$ does not drop rapidly. This is because $\left(\frac{1-\qol}{\qlo}\right)^{n}$
is not large enough and the constant term $1$ in (\ref{eq:23-1})
cannot be neglected. Nevertheless, when $n$ grows large, $\pe$ decreases
much faster in the plot, which is what (\ref{eq:25}) predicted.

An interesting phenomenon can be observed in Fig. \ref{fig:pe}: when
$A_{w}$ grows large, the relative error rate $\pe$ does not approach
$1$, and the line of $\pe$ approaches a plateau as $A_{w}\rightarrow\infty$.
This can be explained from the above results. When $A_{w}\rightarrow\infty$,
the leading terms of $\eta_{0}$ and $\eta_{1}$ in Eq. (\ref{eq:3})
are $|A_{w}|^{2}\sin^{2}n\varphi$ and $|A_{w}|^{2}\cos^{2}n\varphi$
respectively, so
\begin{equation}
\begin{aligned}\lim_{A_{w}\rightarrow\infty}\pe & =\frac{\qlo[n]\cos^{2}n\varphi}{(1-\qol)^{n}\sin^{2}n\varphi+\qlo[n]\cos^{2}n\varphi}\\
 & =\Big(1+\Big(\frac{1-\qol}{\qlo}\Big)^{n}\tan^{2}n\varphi\Big)^{-1}.
\end{aligned}
\label{eq:22}
\end{equation}
This is an upper bound on the relative error rate over all possible
weak values. It can be used to find a suitable $n$ for a given error
rate in the postselection, regardless of the magnitude of $A_{w}$.
Moreover, when $n$ is large (but $n\varphi\ll1$), Eq. (\ref{eq:22})
can be simplified to
\begin{equation}
\lim_{A_{w}\rightarrow\infty}\pe\approx\varphi^{-2}n^{-2}\Big(\frac{1-\qol}{\qlo}\Big)^{-n}.\label{eq:48}
\end{equation}
This implies that the upper bound of the relative error rate $\pe$
can also decay super-exponentially with $n$, which verifies the advantage
of entanglement in suppressing the effect of readout errors.

Note that Eq. (\ref{eq:48}) does not contradict with (\ref{eq:25}).
In (\ref{eq:25}), it is assumed that $n\varphi A_{w}\ll1$, which
requires $A_{w}$ not to be too large, while in (\ref{eq:48}), we
take the limit $A_{w}\rightarrow\infty$. Since the weak value amplification
is a linear approximation theory which works in the regime $n\varphi A_{w}\ll1$,
Eq. (\ref{eq:25}) can be used in practice. And Eq. (\ref{eq:48})
is mainly to provide an upper bound for $\pe$.

\subsection{Loss of correct postselection results\label{sub:Loss-rate}}

In this subsection, we turn to another important effect of readout
errors: the loss of correct postselection results when $\ko$ is misread
as $\kl$. Since the probability of postselecting the $n$ entangled
qubits in the state $\kon$ is $\pon$, and among all postselection
results $\kon$ the proportion of correct states is $(1-q_{0\rightarrow1})^{n}$,
the probability of correct postselection results is
\begin{equation}
p(\mathrm{correct})=\pon(1-q_{0\rightarrow1})^{n}.
\end{equation}
The reduction factor $(1-q_{0\rightarrow1})^{n}$ quantifies the loss
of correct postselection results caused by readout errors in the postselection
measurements. The loss rate for correct postselection is therefore
\begin{equation}
\rl=1-(1-\qol)^{n}.\label{eq:5}
\end{equation}
If the readout error probability $\qol$ is small, so that $n\qol\ll1$,
then
\begin{equation}
\rl\approx n\qol.\label{eq:51}
\end{equation}

Comparing (\ref{eq:25}) and (\ref{eq:51}), one finds that when $n$
increases, the error rate falls but the loss rate grows, and vice
versa. It implies a complementary relation between the error rate
and the loss rate.

This relation for the limiting case $A_{w}\rightarrow\infty$ (which
maximizes the relative error rate $\pe$) can be obtained in the following
way. Suppose $\cdi\sx\di=0$ and $A_{w}\rightarrow\infty$, then by
solving for $n$ from (\ref{eq:5}) and plugging it into (\ref{eq:22}),
we find that
\begin{equation}
\begin{aligned} & \lim_{A_{w}\rightarrow\infty}\pe\\
 & =\bigg(1+(1-\rl)q_{1\rightarrow0}^{-\frac{\log(1-\rl)}{\log(1-q_{0\rightarrow1})}}\tan^{2}\frac{\varphi\log(1-\rl)}{\log(1-q_{0\rightarrow1})}\bigg)^{-1}.
\end{aligned}
\end{equation}
It shows the complementary relation between the relative error rate
and the loss rate due to readout errors when $A_{w}\rightarrow\infty$.

\begin{figure}
\subfloat[$\protect\qlo=\protect\qol=0.05$]{\protect\includegraphics[scale=0.47]{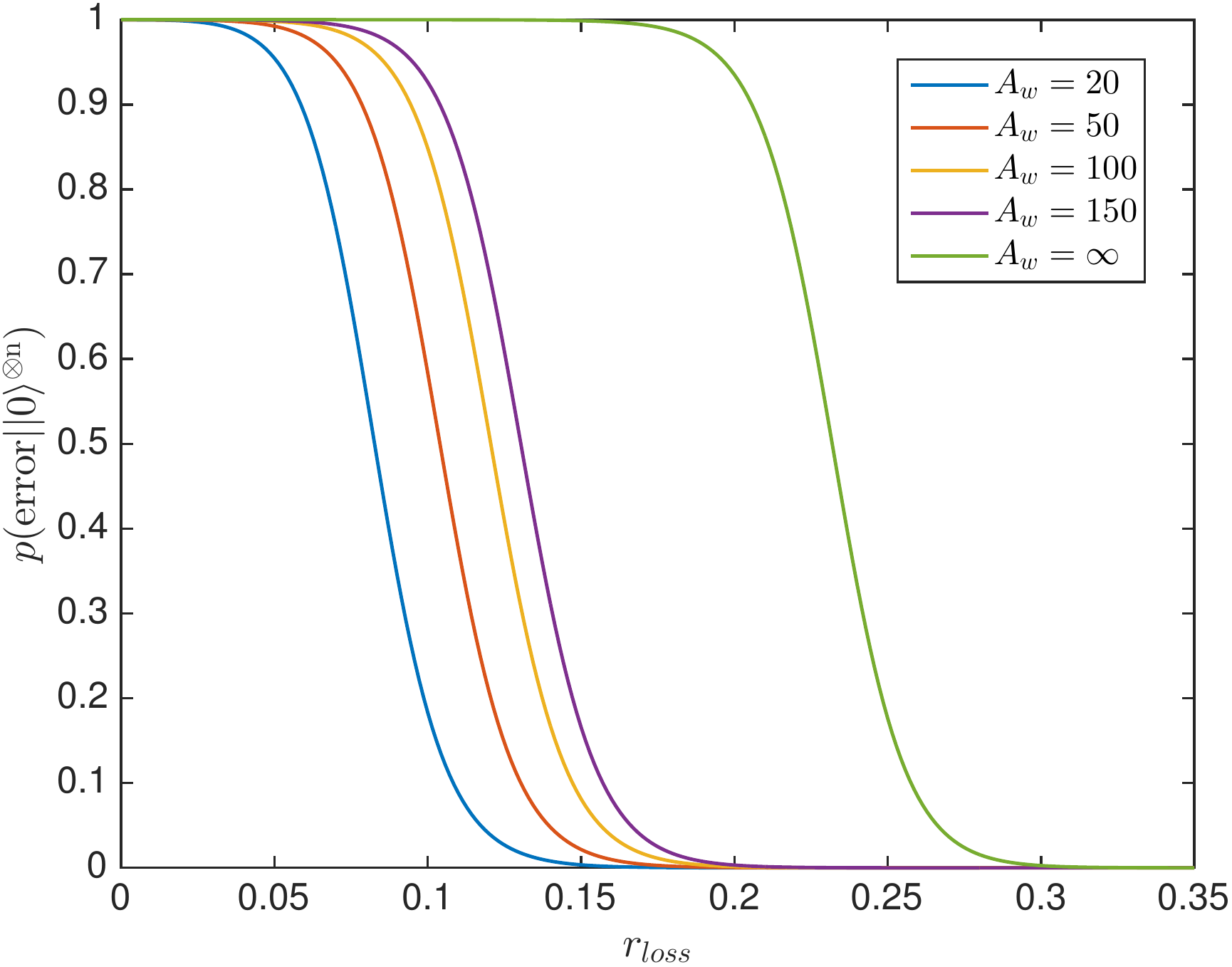}}\\\subfloat[\label{fig:per001}$\protect\qlo=\protect\qol=0.01$]{\protect\includegraphics[scale=0.47]{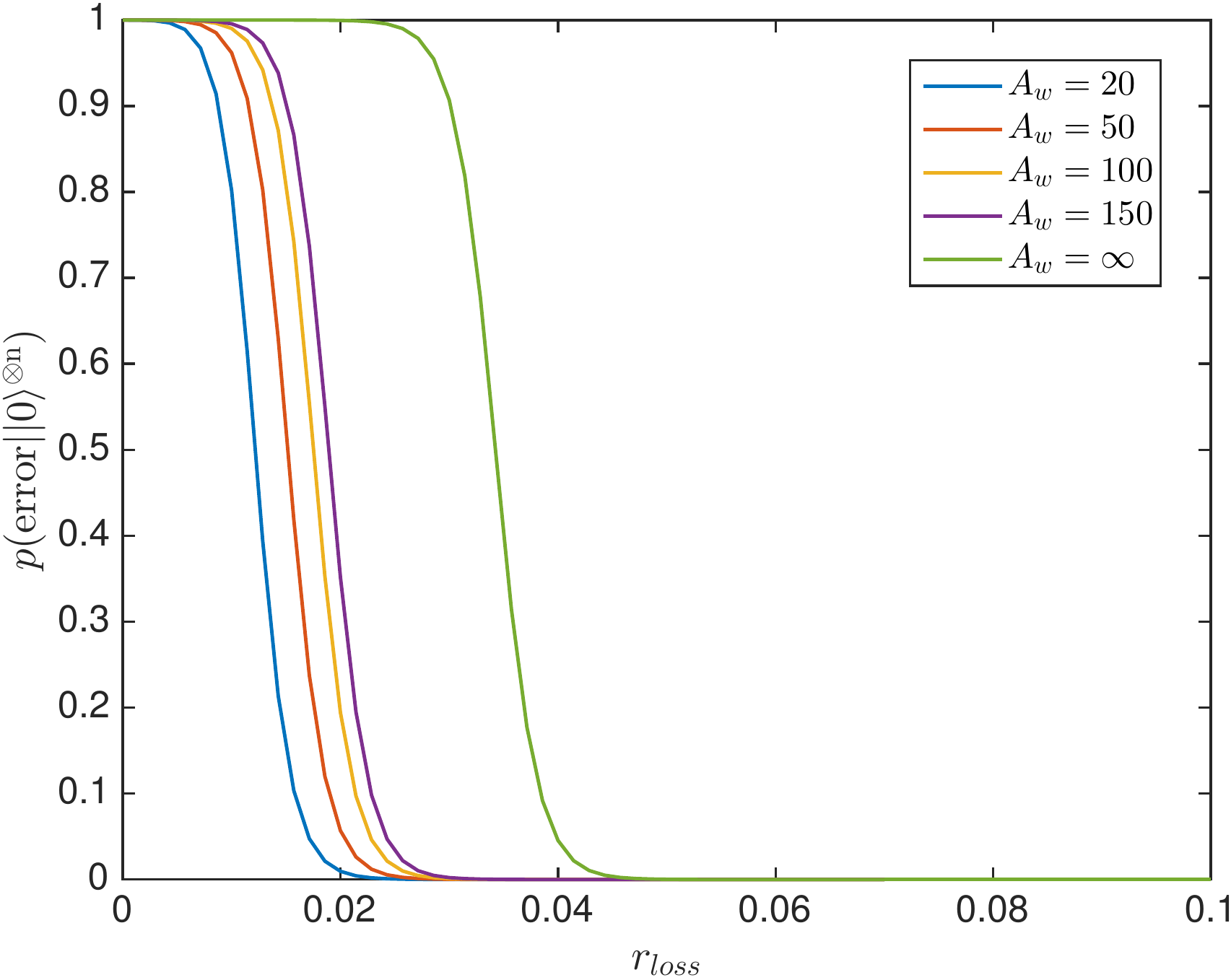}}\protect\caption{\label{fig:per}(Color online) Plot of the complementarity relation
between the maximum relative error probability $p(\mathrm{error}||0\rangle^{\otimes n})$
and the loss rate $\protect\rl$ for different weak values $A_{w}$.
The lines from left to right in each figure are for $A_{w}=20,\,50,\,100,\,150,\,\infty,$
respectively. The $\protect\rl$ axis in (\ref{fig:per001}) is rescaled.}
\end{figure}

In Fig. \ref{fig:per}, the relative error rate $\pe$ is plotted
versus the loss rate $\rl$ for different weak values. As before,
$\qol$ and $\qlo$ are assumed to be equal. The complementary relation
between $\pe$ and $\rl$ can be explicitly observed there.

\subsection{Influence on the measurement result\label{sub:Correction factor}}

In the previous two subsections, we studied the two main effects of
the readout errors: the relative error rate in total postselection
results and the loss rate of correct postselection results. But in
practical applications, what people care about is the final result
of the measurement. So, how significantly do the readout errors affect
the measurement result? And does the entanglement help to suppress
their influence? It is important to make this question clear.

In this subsection, we investigate the modification of the average
measurement result from the pointer qubit when readout errors exist
in the postselection. We will obtain the modified average measurement
result in the presence of readout errors, and show how entanglement
can suppress the influence of readout errors on the measurement result.

Suppose we measure $\sx$ on the pointer qubit after postselecting
the $n$ system qubits, for example. According to $\df[,0]$ and $\df[,1]$
in (\ref{eq:23}) and (\ref{eq:24}), the average shift of the pointer
is
\begin{equation}
\begin{aligned}\delta\langle\sx\rangle_{0} & =-\frac{n\sin2n\varphi\im A_{w}(1-\langle\sx\rangle_{D}^{2})}{\eon},\\
\delta\langle\sx\rangle_{1} & =\frac{n\sin2n\varphi\im A_{w}(1-\langle\sx\rangle_{D}^{2})}{\eln},
\end{aligned}
\label{eq:27}
\end{equation}
for postselection states $\kon$ and $\kln$ respectively, where $\eon$
and $\eln$ are defined in Eq. (\ref{eq:3}). So, taking readout error
into account, the real average result from the pointer qubit when
each system qubit is postselected to $\ko$ is
\begin{equation}
\begin{aligned}\overline{\delta\langle\sx\rangle} & =\frac{\pon(1-\qol)^{n}\delta\langle\sx\rangle_{0}+\pln\qlo[n]\delta\langle\sx\rangle_{1}}{\pon(1-\qol)^{n}+\pln\qlo[n]}\\
 & =-\gn\frac{n\sin2n\varphi\im A_{w}(1-\langle\sx\rangle_{D}^{2})}{\eon},
\end{aligned}
\end{equation}
where
\begin{equation}
\gn=\frac{\pon((1-\qol)^{n}-\qlo[n])}{\pon(1-\qol)^{n}+\pln\qlo[n]}.\label{eq:49}
\end{equation}
Note that $\langle\hat{\sigma}_{x}^{2}\rangle_{D}=1$. Therefore,
$1-\langle\sx\rangle_{D}^{2}=\var(\sx)_{D}$, and thus,
\begin{equation}
\overline{\delta\langle\sx\rangle}=-\gn\frac{n\sin2n\varphi\im A_{w}\var(\sx)_{D}}{\eon}.
\end{equation}

When there is no readout error, the average result from the pointer
qubit is $-n\sin2n\varphi\im A_{w}\var(\sx)_{D}/\eon$, which is approximately
$-2\varphi\im A_{w}\var(\sx)_{D}$ when $nA_{w}\varphi\ll1$. Therefore,
readout errors change the average measurement result of the pointer
qubit by the overall factor $\gn$. Note that
\begin{equation}
\gn=1-\frac{\pe}{\pln},\label{eq:40}
\end{equation}
from Eq. (\ref{eq:4}), which relates the correction factor $\gn$
to the relative error rate $\pe$. 

According to Eq. (\ref{eq:49}), one can deduce that $|\gn|\leq1$,
which means that readout errors reduce the weak value amplification,
but never enhance it. This can be understood intuitively: the postselection
results $\kln$ do not give an amplification of $\varphi$ as the
results $\kon$ do; so when they are mixed with the correct postselection
results $\kon$, the amplification factor will always fall.

\begin{figure}
\subfloat[$\protect\qlo=\protect\qol=0.05$]{\protect\includegraphics[scale=0.47]{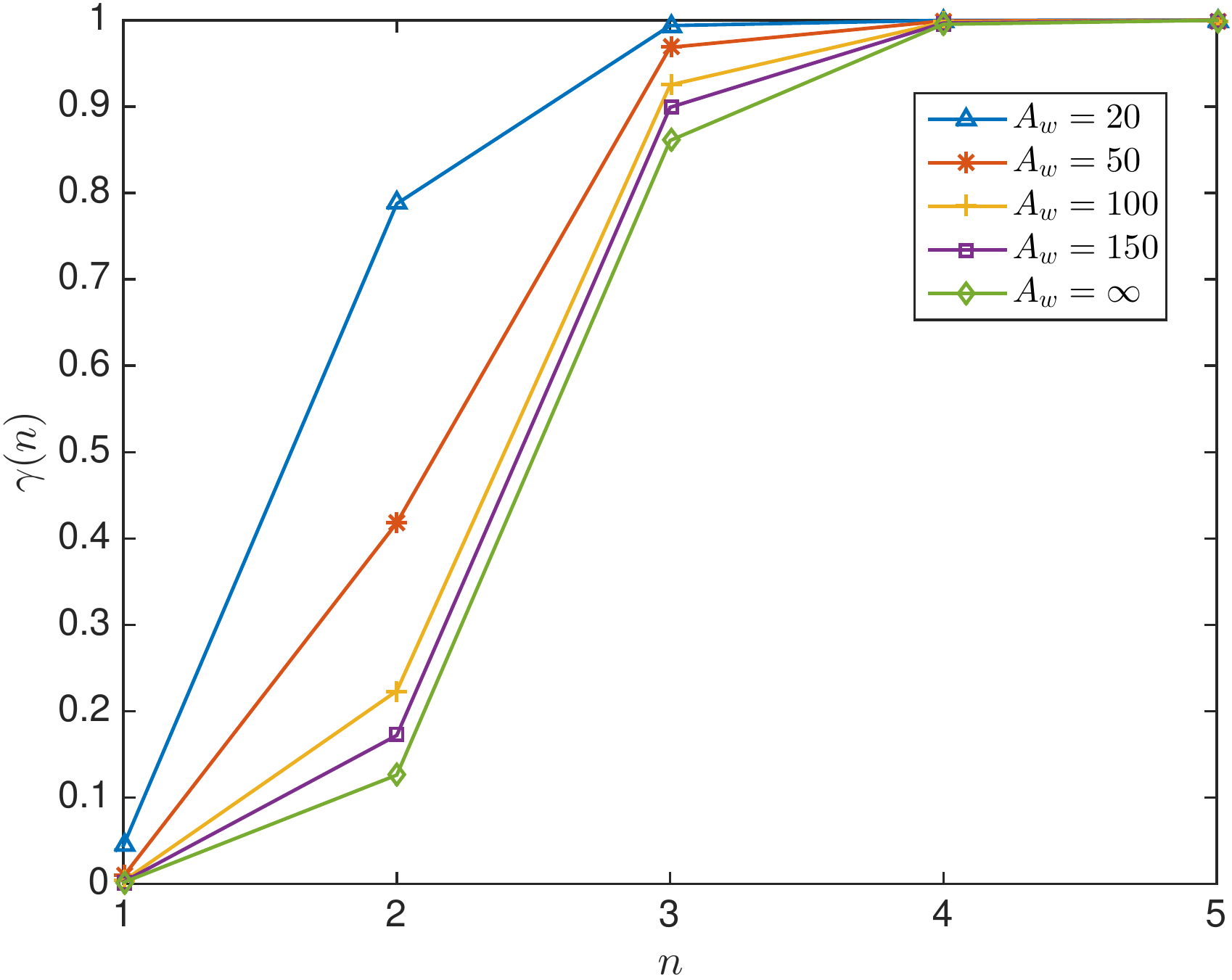}}\\\subfloat[$\protect\qlo=\protect\qol=0.01$]{\protect\includegraphics[scale=0.47]{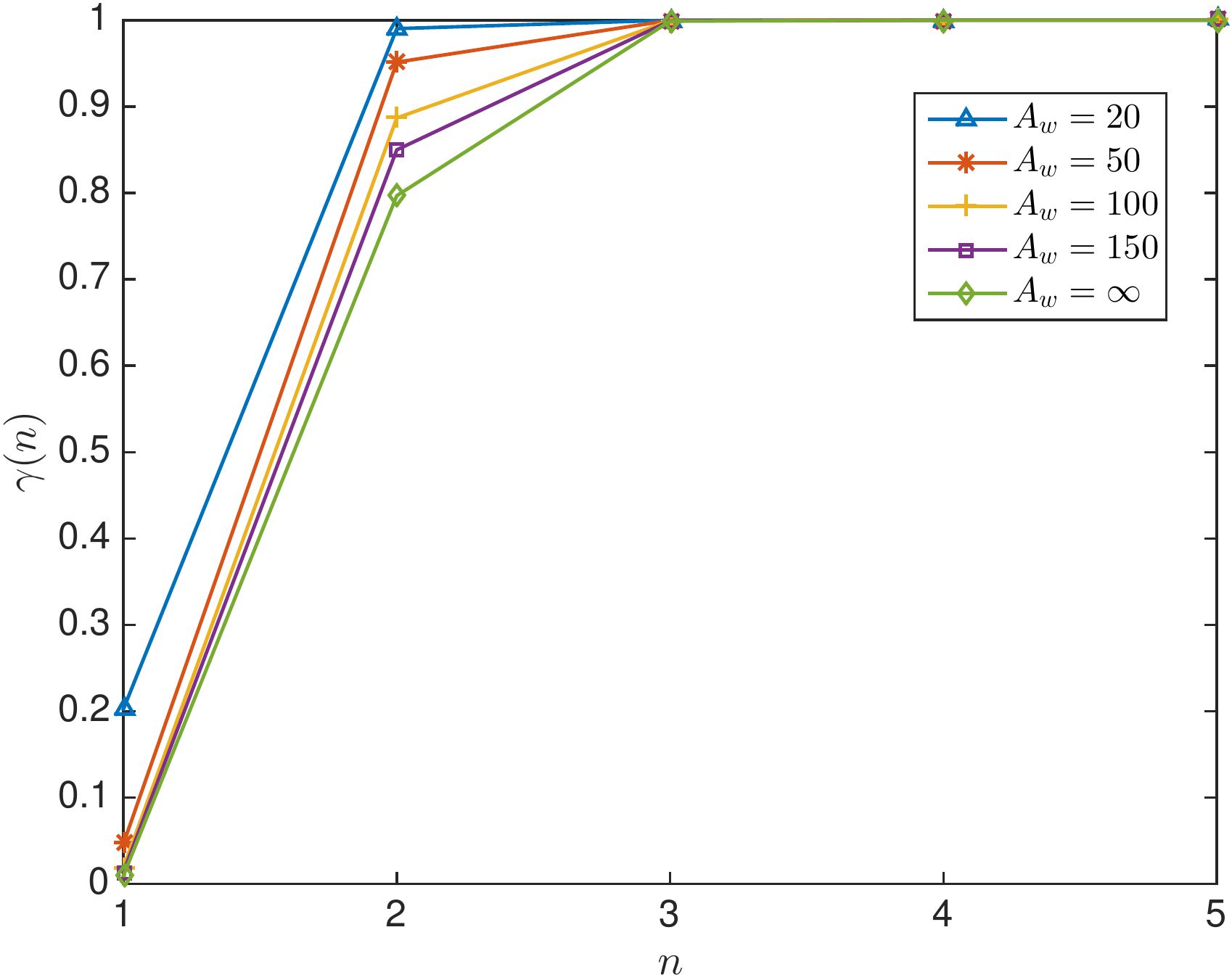}}\protect\caption{\label{fig:gamma}(Color online) This figure plots the correction
factor $\protect\gn$ of the measurement result versus the number
of entangled qubits $n$ for different weak values $A_{w}$. It can
be seen that the correction factor can increase to $1$ very fast
with $n$, which shows the effectiveness of entanglement in protecting
the weak value amplification against readout errors.}
\end{figure}

Notably, since the relative error rate $\pe$ can decrease super-exponentially
fast with the number of entangled qubits $n$ according to (\ref{eq:25}),
and $\pln$ is close to $1$ in the weak value amplification, the
correction factor $\gn$ can then be increased to $1$ very efficiently
with $n$. This is verified by the numerical results in Fig. \ref{fig:gamma},
where the correction factor $\gn$ versus the number of entangled
qubits, $n$, is plotted for different weak values (including $A_{w}=\infty$).
The results again show how entanglement can significantly strengthen
the postselected weak measurement against readout errors.

\subsection{Influence on Fisher information}

In the last subsection, it was shown that readout errors reduce the
average measurement result by the factor $\gamma$ (\ref{eq:40}),
which is linear in the relative error rate $\pe$, and does not depend
on the loss rate $\rl$. This suggests that the error rate in the
postselected events affects the result of the weak measurement, but
the loss of correct postselected events does not. Since entanglement
between the system qubits can dramatically decrease the error rate,
it would be sufficient in this sense to use entanglement to suppress
the effect of readout errors, regardless of the loss of correct postselected
events, the rate of which can increase with the number of entangled
qubits.

This is true for the average measurement result, because it does not
depend on the size of the set of (correct) postselected events. The
loss rate mainly affects the size of that set. However, the precision
of estimating $\varphi$ from the average measurement results does
depend on the loss rate, since the estimation precision of a parameter
generally relies on the size of the sample. 

In this subsection, we study the effect of readout errors on the Fisher
information of estimating $\varphi$ in postselected weak measurements.
We will show that the loss rate indeed affects the Fisher information.
This implies that both the loss rate and the average error must be
suppressed to maintain the precision of the measurement.

To compute the Fisher information, suppose we perform a POVM $\{\he_{1},\cdots,\he_{r}\}$
on the pointer state after postselection. Then , when the $n$ entangled
qubits are postselected in the state $\kon$, and the final pointer
state is $\df[0]$, then the probability to observe the $j$th measurement
outcome is
\begin{equation}
\wo=\avdf 0{\he_{j}}.
\end{equation}
Similarly, when the $n$ entangled qubits are postselected in the
state $\kln$, the final pointer state is $\df[1]$, and the probability
to observe the $j$th outcome is
\begin{equation}
\wl=\avdf 1{\he_{j}}.
\end{equation}

According to Eqs. (\ref{eq:23}) and (\ref{eq:24}), when $\varphi A_{w}\ll1$,
$\df[0]$ and $\df[1]$ can be approximated by
\begin{equation}
\begin{aligned}\df[0] & \approx\exp(-\i\varphi A_{w}\sx)\di,\\
\df[1] & \approx\exp\Big(\i\varphi\frac{n^{2}}{A_{w}^{*}}\sx\Big)\di,
\end{aligned}
\end{equation}
So the probabilities of obtaining the $j$th measurement outcome from
pointer states $\df[0]$ and $\df[1]$ are, respectively,
\begin{equation}
\begin{aligned}\wo & =\frac{\avd{\exp(\i\varphi A_{w}^{*}\sx)\he_{j}\exp(-\i\varphi A_{w}\sx)}}{\avd{\exp(2\varphi\im A_{w}\sx)}},\\
\wl & =\frac{\avd{\exp(-\i\frac{n^{2}}{A_{w}}\varphi\sx)\he_{j}\exp(\i\frac{n^{2}}{A_{w}^{*}}\varphi\sx)}}{\avd{\exp\big(-2\frac{n^{2}}{|A_{w}|^{2}}\varphi\im A_{w}\sx\big)}}.
\end{aligned}
\label{eq:14-2}
\end{equation}
Therefore, the total probability of observing the $j$th outcome from
the final pointer state is
\begin{equation}
h_{j}=\pon\wo(1-\qol)^{n}+\pln\wl\qlo[n],
\end{equation}
where the probability of readout errors has been included.

Since $\varphi\ll1$, we focus on the zeroth order of the Fisher information,
i.e., $\varphi=0$. The Fisher information we acquire from the pointer
state by the POVM $\{\he_{1},\cdots,\he_{r}\}$ is
\begin{equation}
\iv=\sum_{j}\frac{(\pv h_{j})^{2}}{h_{j}}.\label{eq:15-1}
\end{equation}
The term $\pv h_{j}$ can be expanded as
\begin{equation}
\begin{aligned}\pv h_{j} & =(\wo\pv\pon+\pon\pv\wo)(1-\qol)^{n}\\
 & +(\wl\pv\pln+\pln\pv\wl)\qlo[n].
\end{aligned}
\end{equation}

From (\ref{eq:14-2}), we obtain that
\begin{equation}
\begin{aligned}\wl|_{\varphi=0} & =\wo|_{\varphi=0},\\
\pv\wl|_{\varphi=0} & =-\frac{n^{2}}{|A_{w}|^{2}}\pv\wo|_{\varphi=0},
\end{aligned}
\label{eq:44}
\end{equation}
and Eqs. (\ref{eq:2}), (\ref{eq:3}) imply that
\begin{equation}
\begin{aligned}\pln|_{\varphi=0} & =\frac{|A_{w}|^{2}}{n^{2}}\pon|_{\varphi=0},\\
\pv\pln|_{\varphi=0} & =-\pv\pon|_{\varphi=0}.
\end{aligned}
\label{eq:45}
\end{equation}

Plugging these equations into (\ref{eq:15-1}), we get
\begin{equation}
\begin{aligned}\iv & =\frac{n^{2}((1-\qol)^{n}-\qlo[n])^{2}}{n^{2}(1-\qol)^{n}+|A_{w}|^{2}\qlo[n]}\\
 & \times\sum_{j}\frac{(\wo\pv\pon+\pon\pv\wo)^{2}}{\pon\wo}.
\end{aligned}
\label{eq:46}
\end{equation}
When there are no readout errors, $\qlo=\qol=0$, so the Fisher information
is
\begin{equation}
\ivo=\sum_{j}\frac{(\wo\pv\pon+\pon\pv\wo)^{2}}{\pon\wo}.\label{eq:47}
\end{equation}
Therefore, the Fisher information modified by the readout errors can
be written as
\begin{equation}
\ivo=f(n)\ivo,
\end{equation}
where
\begin{equation}
f(n)=\frac{n^{2}((1-\qol)^{n}-\qlo[n])^{2}}{n^{2}(1-\qol)^{n}+|A_{w}|^{2}\qlo[n]}.\label{eq:42}
\end{equation}

The factor $f(n)$ quantifies the effect of readout errors on the
Fisher information. Note that $f(n)<1$ when $\qlo\neq0,\,\qol\neq0$,
so readout errors always reduce the Fisher information of the weak
measurement, and never increase it. Eq. (\ref{eq:42}) also shows
the role of the number of entangled qubits $n$ and the weak value
$A_{w}$ on the extent to which the readout errors can reduce the
Fisher information. 

\begin{figure}
\subfloat[$\protect\qlo=\protect\qol=0.05$]{\protect\includegraphics[scale=0.47]{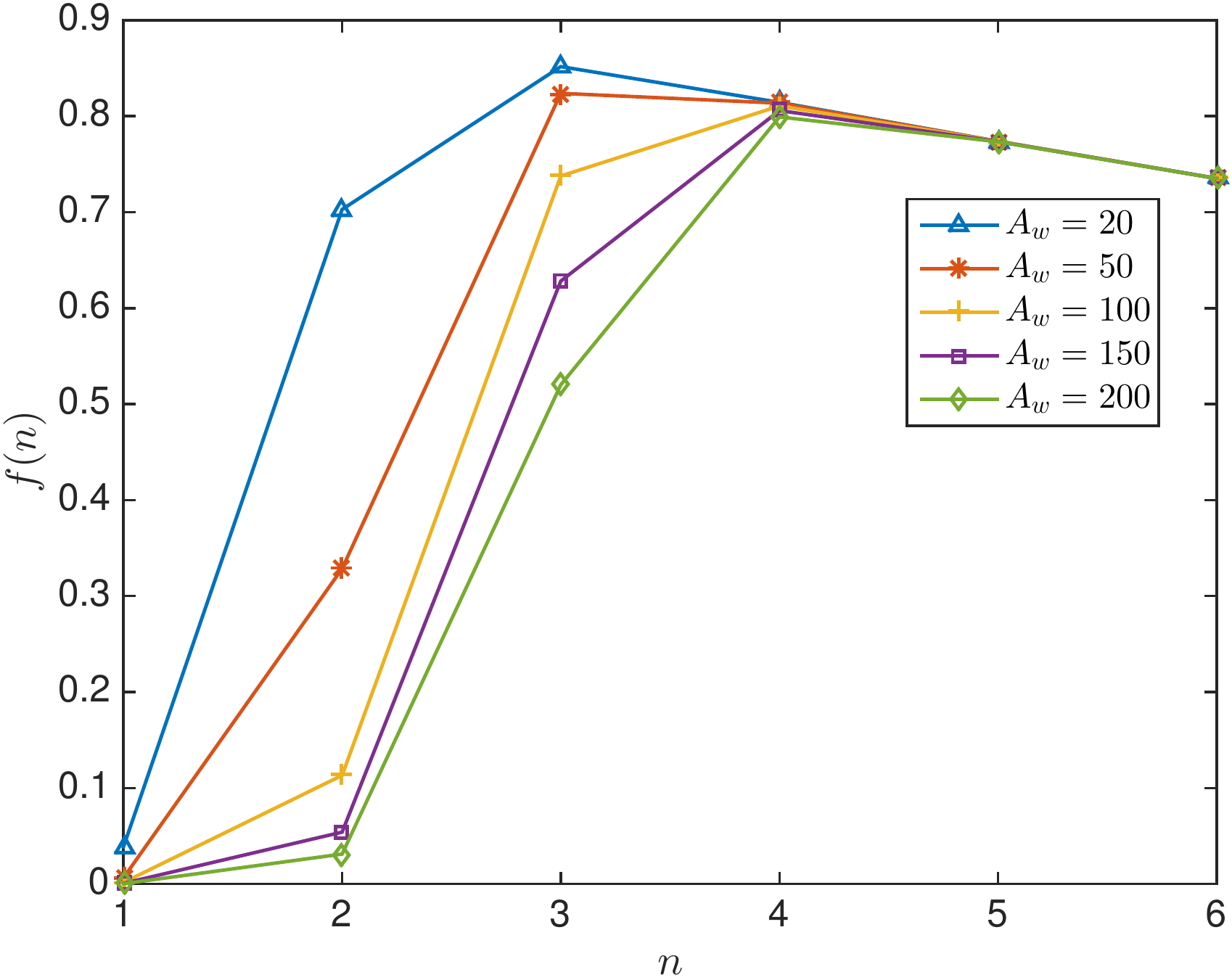}}\\\subfloat[$\protect\qlo=\protect\qol=0.01$]{\protect\includegraphics[scale=0.47]{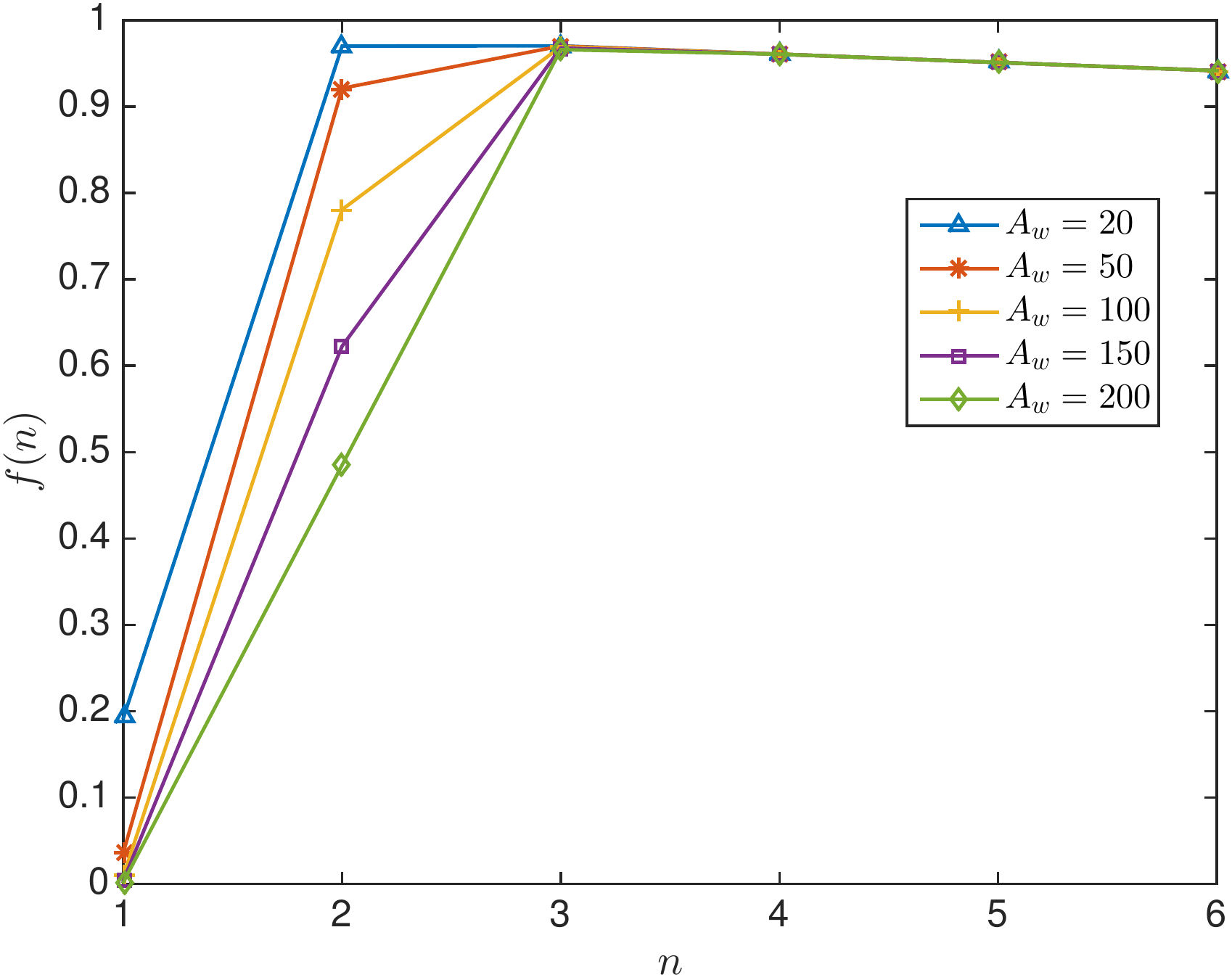}}\protect\caption{(Color online) This figure plots the modification factor $f(n)$ of
the Fisher information versus the number of entangled qubits $n$
in the presence of readout errors, but without majority voting, for
different weak values. When $n$ is small, the Fisher information
can increase with $n$, because the entanglement eliminates some errors
in the postselection results and raises the proportion of correct
postselected states that have higher Fisher information. However,
when $n$ becomes larger, the Fisher information starts to fall, since
the loss rate of correct postselected states dramatically increases
with $n$ in this case.\label{fig:no majority voting}}
\end{figure}

Fig. \ref{fig:no majority voting} plots how the Fisher information
changes as $n$ increases. It shows that entanglement can recover
some lost Fisher information by raising the average shift, but the
Fisher information is still reduced by loss. As we will see in the
next subsection, when entanglement is combined with a majority voting
scheme, more Fisher information can be recovered.

Before concluding this subsection, we want to point out the relation
between the modification factor $f(n)$ for Fisher information and
the loss rate of correct postselected events $\rl$. Generally, the
probability of a single readout error is low, so $\qlo[n]\rightarrow0$
when $n$ is not small. And according to Eq. (\ref{eq:5}), $(1-\qol)^{n}=1-\rl$.
Hence, from (\ref{eq:42}) we immediately have
\begin{equation}
f(n)\approx1-\rl.\label{eq:43}
\end{equation}
This explicitly shows the relation between $f(n)$ and the loss rate
$\rl$, and verifies that the loss of correct postselected events
does indeed cause a reduction in the Fisher information of the weak
measurement.

The relation (\ref{eq:43}) can also be understood in a more intuitive
way: the Fisher information is proportional to the size of the sample
that is used for parameter estimation, and the proportion of correct
postselection results in the whole set of postselected events is $1-\rl$,
so the Fisher information is reduced by $\rl$, as indicated by (\ref{eq:43}).
This suggests the necessity of suppressing the loss rate as well as
the relative error rate.

\subsection{Majority voting scheme for recovering Fisher information}

As the loss of correct postselection results can be detrimental to
the Fisher information of the weak measurement, it is necessary to
eliminate or suppress the loss. In this subsection, we introduce a
majority voting scheme on the postselection results to decrease the
loss rate of the correct postselection results and increase the effective
Fisher information of the weak measurement.

The idea comes from a simple observation on Eq. (\ref{eq:21}): the
true postselected states of the $n$ system qubits are correlated,
and should either be all $\ko$ or all $\kl$. So when readout errors
occur, it is still possible to determine whether the postselection
is successful or not with high probability from the majority of the
observed states of the $n$ qubits. If one observes more $\ko$'s
than $\kl$'s, it is more likely that the $n$ system qubits are postselected
to $\kon$. And vice versa. This is the majority voting scheme.

In this subsection, we will examine this scheme carefully, and show
that it can effectively suppress the loss of the correct postselection
results, and recover the lost Fisher information of the weak measurement.

Suppose at most $k$ readout errors are allowed in the postselected
results of a batch of $n$ qubits (i.e. the number of observed $\ko$'s
or $\kl$'s, whichever is lesser, is no more than some threshold $k$),
and assume that the readout errors are independent of each other.
The loss rate of the correct postselected states in this case becomes
\begin{equation}
\rlp=1-\sk(1-\qol)^{n-j}\qlo[j],\label{eq:46-1}
\end{equation}
which is obviously lower than Eq. (\ref{eq:5}), so more correct postselected
events are retained by the majority voting scheme.

Now, let us study the Fisher information $\iv$ when majority voting
is used. The total probability of postselecting the $n$ qubits in
the state $\kon$ is
\begin{equation}
\begin{aligned}\hk & =\pon\wo\sk(1-\qol)^{n-j}\qlo[j]\\
 & +\pln\wl\sk\qlo[n-j](1-\qol)^{j}.
\end{aligned}
\end{equation}
Note that in the expansion of $\partial_{\varphi}\hk$ in this case,
Eqs. (\ref{eq:44}) and (\ref{eq:45}) are unchanged. Therefore, the
new Fisher information with the majority voting scheme can be derived
by plugging the following replacement into Eq. (\ref{eq:46}):
\begin{equation}
\begin{aligned}(1-\qol)^{n} & \longrightarrow\sk(1-\qol)^{n-j}\qlo[j],\\
\qlo[n] & \longrightarrow\sk\qlo[n-j](1-\qol)^{j}.
\end{aligned}
\end{equation}
The result is
\begin{equation}
\iv=f(n,k)\ivo,\label{eq:5-1}
\end{equation}
where the factor $f(n,k)$ is
\begin{equation}
\begin{aligned} & f(n,k)=\\
 & \frac{n^{2}\bigg({\displaystyle \sk\left((1-\qol)^{n-j}\qlo[j]-\qlo[n-j](1-\qol)^{j}\right)}\bigg)^{2}}{{\displaystyle \sk\left(n^{2}(1-\qol)^{n-j}\qlo[j]+|A_{w}|^{2}\qlo[n-j](1-\qol)^{j}\right)}},
\end{aligned}
\end{equation}
and $\ivo$ is still the original Fisher information without readout
error, the same as Eq. (\ref{eq:47}).

Just like the modification factor $f(n)$ in the case of readout errors
without majority voting, the overall factor $f(n,k)$ in Eq. (\ref{eq:5-1})
determines the total change in the Fisher information due to readout
errors in the presence of majority voting. When the probability of
a readout error is sufficiently low, the $f(n,k)$ can be roughly
approximated by
\begin{equation}
f(n,k)\approx1-\rlp,
\end{equation}
where $\rlp$ is given in (\ref{eq:46-1}), so it is only to be expected
that part of the Fisher information will be recovered, since majority
voting can retain some originally lost postselected events.

To see how efficiently the majority voting scheme can work for protecting
the Fisher information against the readout errors, we study the factor
$f(n,k)$ in detail numerically.

\begin{figure}
\subfloat[\label{fig:fk-a}$\protect\qlo=\protect\qol=0.05$]{\protect\includegraphics[scale=0.47]{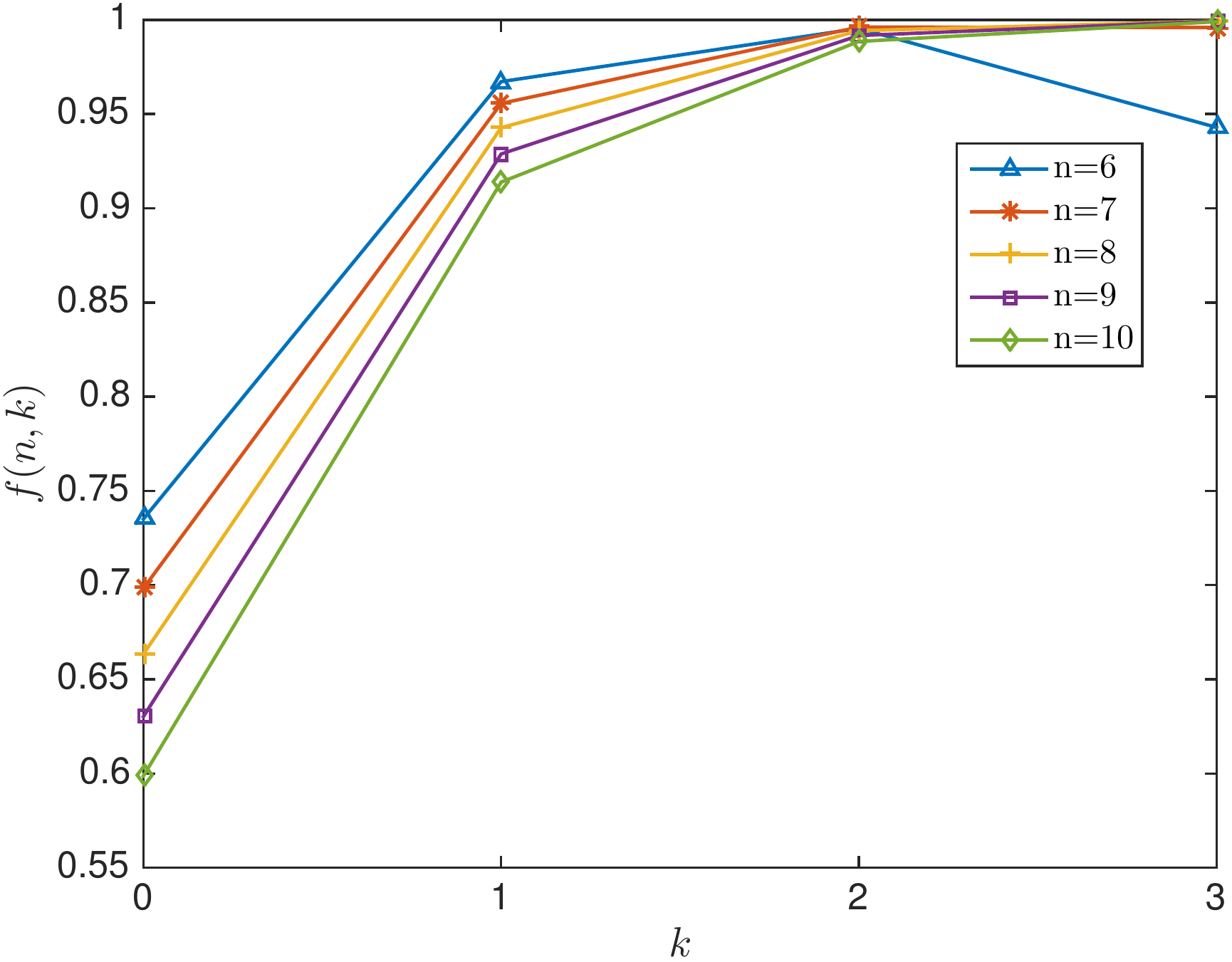}

}

\subfloat[\label{fig:fk-b}$\protect\qlo=\protect\qol=0.01$]{\protect\includegraphics[scale=0.47]{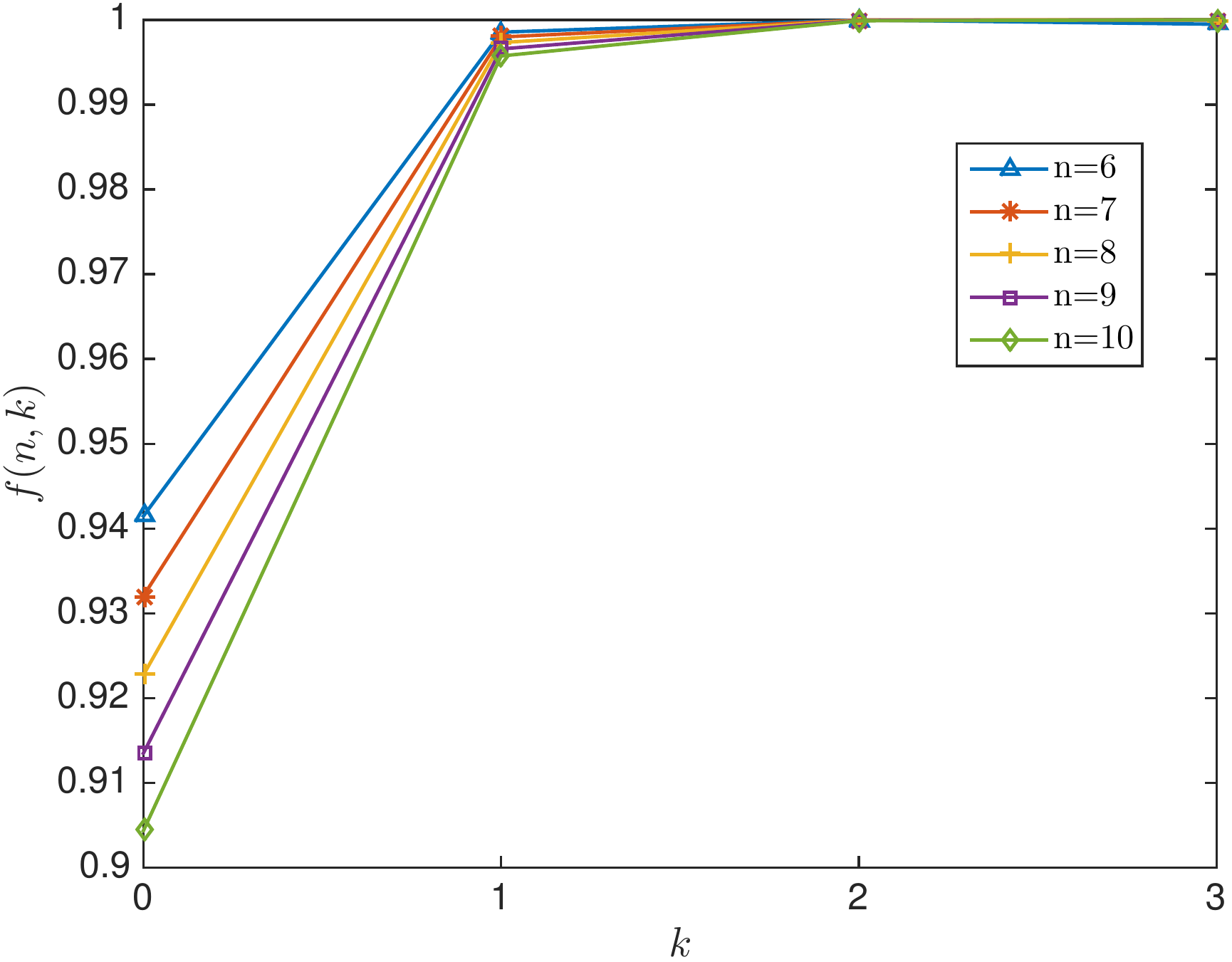}}\protect\caption{(Color online) This figure plots the modification factor $f(n,k)$
of the Fisher information versus the allowed number of readout errors
$k$ for different $n$ when the majority voting scheme is used. The
number of allowed readout errors, $k$, is from $0$ to $3$. The
weak value is $A_{w}=30$. The figure shows that when $k$ climbs
to half of $n$, $f(n,k)$ can be very close to $1$. It implies that
with a proper number of readout errors allowed in postselecting the
$n$ qubits, the majority voting strategy can recover almost all the
Fisher information lost by the readout errors.\label{fig:fk}}
\end{figure}

Fig. \ref{fig:fk} plots the factor $f(n,k)$ versus $k$ for different
$n$. The line $k=0$ corresponds to the case without majority voting.
The figure shows the modification factor $f(n,k)$ has a dramatic
increase from $k=0$ to $k=1$ (and larger $k$), and can almost reach
$1$ with proper $k$. This implies that with the majority voting
scheme, the loss of correct postselection results can be almost completely
suppressed, and nearly all of the lost Fisher information can be recovered. 

This is a remarkable result. By contrast, when no majority voting
scheme is used, the lost Fisher information can only be partially
recovered by the entanglement, as indicated by Fig. \ref{fig:no majority voting}.
This verifies the effectiveness of the majority voting scheme in protecting
Fisher information against readout errors.

A notable point in Fig. \ref{fig:fk-a} is that for $n=6$, when $k=3$,
there is a small drop in $f(n,k)$. This is because when $n$ is even
and $k=n/2$, if the probability of readout errors is not small, a
large fraction of the failed postselections will be identified as
successful ones, and the failed postselections contain much lower
Fisher information than the successful ones. So the Fisher will fall
in this case. However, if the probability of readout errors is sufficiently
low, the fraction of misidentified postselections will be very small,
then the Fisher information will not drop. Fig. \ref{fig:fk-b} shows
the latter case. 

It is also worth mentioning that according to Eq. (\ref{eq:5-1}),
the effect of readout errors with the majority voting scheme on the
Fisher information is an overall reduction by the factor $f(n,k)$,
so when the original Fisher information $\ivo$ is maximized, the
reduced Fisher information $\iv$ is also maximized. This implies
that the optimal measurement to maximize the original Fisher information
will still be the optimal when readout errors exist and the majority
voting scheme is used. So the optimal measurement on the pointer qubit
does not need to change in the presence of readout errors. This may
be convenient for practical applications.

\subsection{Effect of majority voting scheme on the measurement result}

In the last subsection, we showed that the majority voting scheme
can efficiently recover almost all of the Fisher information lost
by readout errors. A separate question is how the measurement result
is affected by the majority voting scheme. In this subsection, we
will investigate this problem in detail. We still measure $\sx$ on
the pointer qubit after postselecting the $n$ system qubits, similar
to Sec. \ref{sub:Correction factor}.

According to Eq. (\ref{eq:27}), if we allow at most $k$ errors in
the postselection result of the $n$ system qubits, the real average
result from the pointer qubit after the postselection is
\begin{equation}
\begin{aligned}\overline{\delta\langle\sx\rangle}\\
= & \bigg(\sk\Big(\pon(1-\qol)^{n-j}\qlo[j]\delta\langle\sx\rangle_{0}\\
 & {\displaystyle +\pln(1-\qol)^{j}\qlo[n-j]\delta\langle\sx\rangle_{1}\Big)}\bigg)\Big/\\
 & \bigg(\sk\Big(\pon(1-\qol)^{n-j}\qlo[j]\\
 & +\pln(1-\qol)^{j}\qlo[n-j]\Big)\bigg)\\
= & -\gnk\frac{n\sin2n\varphi\im A_{w}\var(\sx)_{D}}{\eon},
\end{aligned}
\end{equation}
where we have used $(1-\langle\sx\rangle_{D}^{2})=\var(\sx)_{D}$,
and
\begin{equation}
\begin{aligned} & \gnk=\\
 & \frac{\pon{\displaystyle \sk\Big((1-\qol)^{n-j}\qlo[j]-(1-\qol)^{j}\qlo[n-j]\Big)}}{{\displaystyle \sk\Big(\pon(1-\qol)^{n-j}\qlo[j]+\pln(1-\qol)^{j}\qlo[n-j]\Big)}}.
\end{aligned}
\label{eq:50}
\end{equation}

Similar to the case without majority voting, $\gnk$ is the correction
factor of the average measurement result, using the majority voting
scheme with at most $k$ errors allowed. From Eq. (\ref{eq:50}) one
can deduce that $|\gnk|\leq1$, so the readout errors still reduce
the weak value amplification even when majority voting is used.

\begin{figure}
\subfloat[\label{fig:gnk-a}$\protect\qlo=\protect\qol=0.05$]{\protect\includegraphics[scale=0.47]{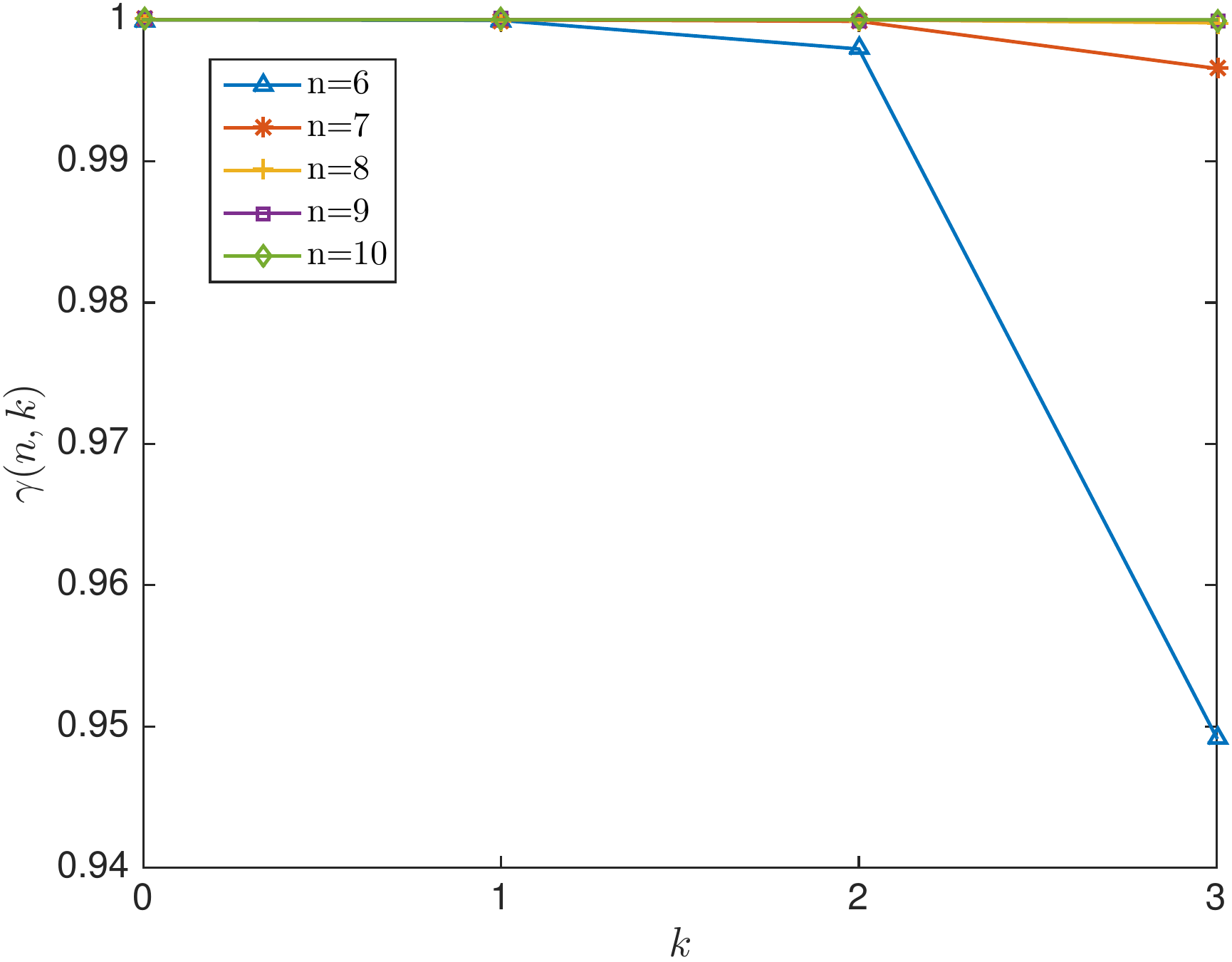}

}

\subfloat[\label{fig:gnk-b}$\protect\qlo=\protect\qol=0.01$]{\protect\includegraphics[scale=0.47]{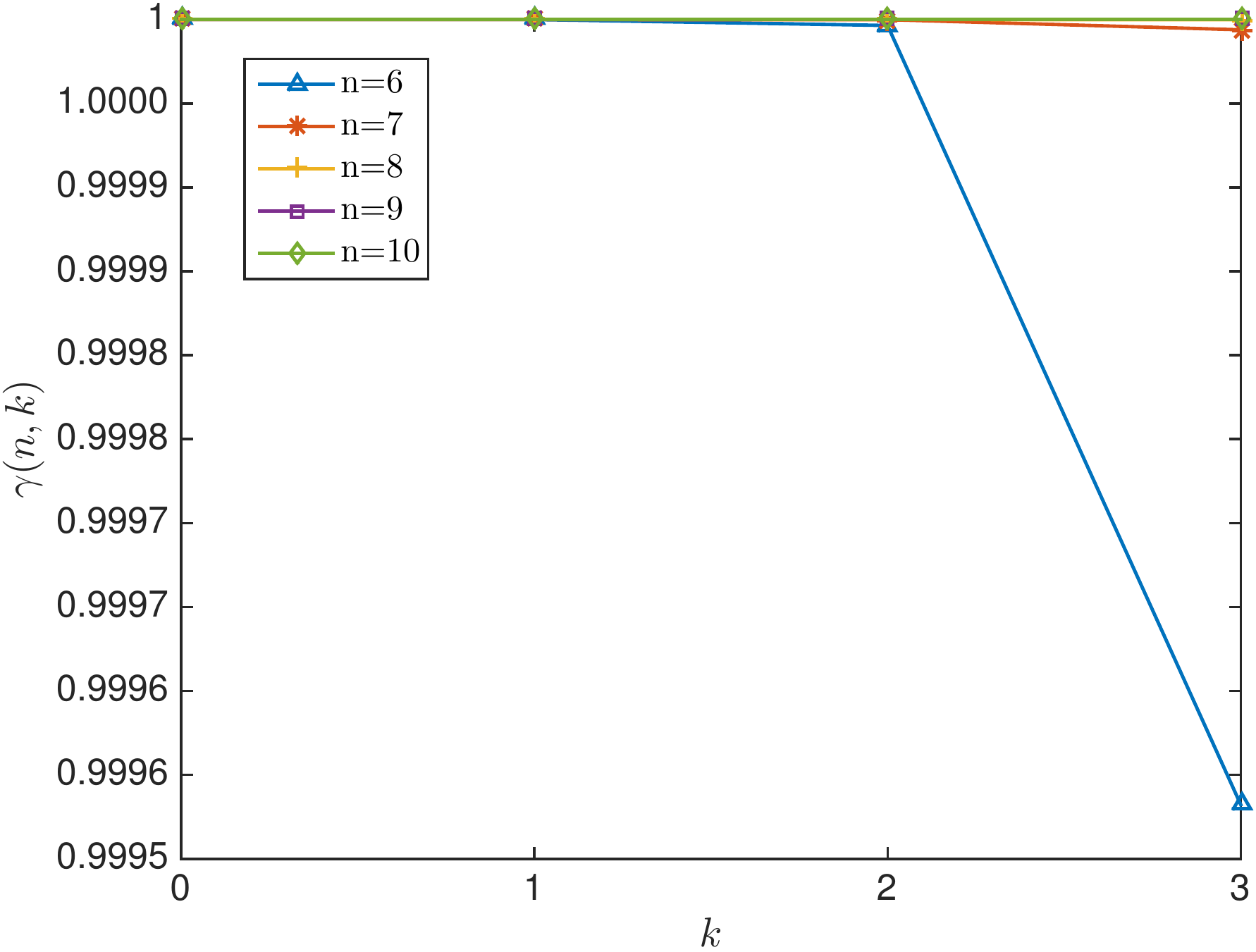}}\protect\caption{(Color online) This figure plots the correction factor $\protect\gnk$
of the measurement result versus the allowed number of readout errors
$k$ for different $n$, when the majority voting scheme is employed.
The weak value is $A_{w}=30$. The lines for $n=8,9,10$ almost overlap,
since they are very close to each other. The figure shows that when
$k$ increases, a small drop in the measurement result may occur,
due to more errors being induced by the majority voting scheme. \label{fig:gnk}}
\end{figure}

Fig. \ref{fig:gnk} plots the factor $\gnk$ versus $k$ for different
$n$. When $k$ increases, the amplification of the measurement result
has a small drop. The reason is similar to that for the drop in the
Fisher information in Fig. \ref{fig:fk-a}. That is, when $k$ increases,
more errors are allowed by the majority voting strategy, and erroneous
postselections correspond to much lower weak values than correct postselections,
so the amplification factor is reduced.

The results of this section and the last section suggest that there
is a balance between the number of entangled qubits $n$ and the allowed
number of readout errors $k$ in the majority voting scheme. On the
one hand, to effectively restore the lost Fisher information, $k$
should not be too small; otherwise, the majority voting scheme cannot
recover most of the lost successful postselections, and a considerable
part of Fisher information will still be discarded. On the other hand,
if $k$ is too large, there will be too many wrong postselection results
mixed into the correction postselections, which will decrease both
the Fisher information and the weak value amplification factor. Therefore,
for a given number of entangled qubits $n$, one needs to find a suitable
number of allowed readout errors $k$, so that the effects of readout
errors can be effectively suppressed.

\section{Summary}

In this paper, we studied the optimization of postselected weak measurements
to improve the performance of weak value amplification. This problem
is approached in two ways: one is to maximize the postselection probability
with a fixed weak value, which aims to improve the usage of resources;
the other is to maximize the weak value for a given postselection
probability, which aims to enhance the amplification ability of weak
measurements. 

We found that both of these can be significantly increased by using
entangled systems, which results in that the Fisher information of
the measurement can also be increased, and can approximately saturate
the Heisenberg limit. Based on this, we proposed a protocol for entanglement-assisted
weak measurement. We provided the optimal choice of initial state
and postselection of the system for this protocol, and illustrated
it by a qubit example with simple quantum circuits.

Furthermore, we considered the influence of readout errors on the
protocol. Readout errors are more harmful to postselected weak measurements
than to other quantum measurements, since even a small rate of readout
errors can give rise to severe disturbance in a postselected weak
measurement when the postselection probability is low. So it is particularly
necessary to consider the effect of readout errors in postselected
weak measurements. 

There are two major problems resulting from readout errors. One is
that an error will occur in the measurement result; the other is that
the Fisher information of the measurement will decrease. We found
that entanglement between the systems can eliminate the error in the
measurement result very efficiently, so the first problem can be solved.
Moreover, entanglement can also retrieve some of the lost Fisher information.
To further suppress the loss of Fisher information, we introduced
a majority voting strategy, and showed that with this strategy, almost
no Fisher information will be lost.

Postselected weak measurement is a useful scheme to measure tiny physical
effects, and how to exploit quantum resources to overcome its low
efficiency and improve the sensitivity is of great interest in practical
applications. It is worth mentioning that recently squeezing was also
found useful in increasing the SNR of postselected weak measurement
\cite{Pang2014b}. We hope that our work will help to deepen the understanding
of this innovative measurement protocol, and extend it to broader
applications.
\begin{acknowledgments}
We thank Justin Dressel for much useful discussion. This research
was supported by the ARO MURI under Grant No. W911NF-11-1-0268.
\end{acknowledgments}

\bibliographystyle{apsrev4-1}
\bibliography{EWVA}

\end{document}